\documentclass[unsortedaddress,superscriptaddress,prb,onecolumn,12pt,aps]{revtex4-2}

\usepackage[T1]{fontenc}
\usepackage{lmodern}
\usepackage[english]{babel}
\usepackage{microtype}

\usepackage{bm}
\usepackage{upgreek}
\usepackage{bbold}
\usepackage{braket}
\usepackage{wasysym}
\usepackage{siunitx}

\usepackage{graphicx}
\usepackage[table]{xcolor}
\usepackage{booktabs}
\usepackage{dcolumn}
\usepackage[para]{threeparttable}
\usepackage{wrapfig}

\usepackage{orcidlink}

\usepackage{soul}
\usepackage[normalem]{ulem}
\usepackage{setspace}
\usepackage{parskip}

\usepackage{caption}
\DeclareCaptionJustification{block}{%
  \leftskip=0pt \rightskip=0pt \parfillskip=0pt plus 1fil \parindent=0pt}
\makeatletter
\newif\if@restonecol
\makeatother
\usepackage{tocloft}
\usepackage{multibib}
\newcites{SM}{References}

\usepackage{hyperref}
\hypersetup{colorlinks=true, allcolors=blue}

\makeatletter
\AtBeginDocument{%
  \@ifundefined{r@LastBibItem}%
    {\@namedef{r@LastBibItem}{{99}{}{}{}{}}}{}%
}
\makeatother

\newcolumntype{s}{p{2.5cm}}
\arrayrulecolor{black}

\DeclareGraphicsExtensions{.pdf}

\newcounter{suppnote}
\newcounter{suppfig}
\newcounter{supptab}

\newcommand{\suppnote}[2]{%
    \par\bigskip
    \refstepcounter{suppnote}%
    \noindent\textbf{Supplementary Note \thesuppnote}\quad #1
    \addcontentsline{toc}{figure}%
        {\textbf{Supplementary Note \thesuppnote}\quad #1}%
    \par\medskip
    #2%
    \par\bigskip
}
\newcommand{\suppfig}[2]{%
    \par\medskip
    \refstepcounter{suppfig}%
    \addcontentsline{toc}{figure}%
        {\textbf{Supplementary Figure \thesuppfig}\quad #1}%
    #2%
    \par\bigskip
}
\newcommand{\supptab}[2]{%
    \par\medskip
    \refstepcounter{supptab}%
    \addcontentsline{toc}{figure}%
        {\textbf{Supplementary Table \thesupptab}\quad #1}%
    #2%
    \par\bigskip
}

\begin{document}
\pagestyle{plain}
\addtocontents{toc}{\protect\setcounter{tocdepth}{-1}}
\title{Atomic-scale assembly of sulfur vacancy complexes\\
for tailored in-gap states in monolayer WS$_{2}$}

\author{Daniel Jansen\,\orcidlink{0009-0002-0763-5853}}
\email{jansen@ph2.uni-koeln.de}
\affiliation{Institute of Physics II, University of Cologne, Z\"{u}lpicher Stra\ss e 77, 50937 Cologne, Germany}
\author{Guangyao Miao\,\orcidlink{0000-0003-3776-5851}}
\affiliation{Institute of Physics II, University of Cologne, Z\"{u}lpicher Stra\ss e 77, 50937 Cologne, Germany}
\author{Jan-Heinrich Keienburg\,\orcidlink{0009-0009-3021-0045}}
\affiliation{Institute of Physics II, University of Cologne, Z\"{u}lpicher Stra\ss e 77, 50937 Cologne, Germany}
\author{Jeison Fischer\,\orcidlink{0000-0003-3538-1926}}
\affiliation{Institute of Physics II, University of Cologne, Z\"{u}lpicher Stra\ss e 77, 50937 Cologne, Germany}
\author{Thomas Michely\,\orcidlink{0000-0003-1657-1847}}
\affiliation{Institute of Physics II, University of Cologne, Z\"{u}lpicher Stra\ss e 77, 50937 Cologne, Germany}
\author{Hannu-Pekka Komsa\,\orcidlink{0000-0002-0970-0957}}
\affiliation{Faculty of Information Technology and Electrical Engineering, University of Oulu, Pentti Kaiteran katu 1, 90014 Oulu, Finland}
\author{Wouter Jolie\,\orcidlink{0000-0001-9319-3427}}
\email{wjolie@ph2.uni-koeln.de}
\affiliation{Institute of Physics II, University of Cologne, Z\"{u}lpicher Stra\ss e 77, 50937 Cologne, Germany}
\affiliation{Institute for Integrated Natural Sciences, Physics Department, University of Koblenz, Universit\"{a}tsstra\ss e 1, 56070 Koblenz, Germany}

\date{\today}
\begin{abstract}
\textbf{Atomic-scale defects in two-dimensional semiconductors are quasi-isolated quantum systems with promise for emerging technologies. Although their potential as spin qubits, quantum sensors and single-photon emitters has been demonstrated experimentally, creating them deterministically and identifying them chemically remains challenging, which limits control over their electronic states. In this work we assemble sulfur vacancies in monolayer $\mathrm{WS_{2}}$ into complexes of different sizes and geometries using the tip of a scanning tunneling microscope. We study their electronic properties by a combination of scanning tunneling spectroscopy and density functional theory, revealing the intricate influences of geometry, underlying point group symmetries and spin-orbit coupling on hybridized states. Among the realized sulfur vacancy complexes, we characterize two configurations that are candidates for bright single photon emitters. Our findings demonstrate an atomic-scale method for engineering quantum states within the band gap of two-dimensional semiconductors.}
\end{abstract}

\maketitle
\thispagestyle{plain}

\section{Introduction}

Point defects in monolayers of semiconducting transition-metal dichalcogenides (TMDs) have a wide range of properties relevant to the emerging field of quantum nanotechnology~\cite{wolfowicz2021quantum, ping2021design, turunen2022photonics, fang2026towards}. 
Their surface accessibility is envisioned to allow creation and manipulation of tailored hybrid orbitals in the band gap of their host. Optical transitions involving these localized orbitals can give rise to antibunched single-photon emission (SPE), first observed in 2D semiconductors more than a decade ago~\cite{he2015single, srivastava2015optical, tonndorf2015single, chakraborty2015quantum}.

Although SPE of defects in semiconducting TMDs is experimentally well established in far-field optical studies, the nature of the defects remains strongly debated as these measurements lack atomic resolution. Single chalcogen vacancies have long been the prime candidates for the narrow photoluminescence (PL) emission lines observed experimentally~\cite{klein2019site, mitterreiter2021role, hoetger2023spin, hoetger2023photovoltage}, which were shown to exhibit SPE characteristics~\cite{barthelmi2020atomistic, klein2021engineering, barthelmi2026emission}. However, near field optical measurements probing photon emission from the tunnel junction of a scanning tunneling microscope (STM) only observed electroluminescence~\cite{schuler2020electrical} with no exciton emission lines emerging at sulfur vacancy sites~\cite{huberich2025exciton}, questioning single chalcogen vacancies as SPE sources. This observation is in line with a weak optical transition dipole calculated for single sulfur vacancies in $\mathrm{WS_{2}}$~\cite{sun2024unveiling}, which only becomes comparable to the direct band-to-band transitions of the pristine 2D semiconductor under application of strain~\cite{hernandez2022strain}. 

Recently, complexes comprising two or more chalcogen vacancies have been proposed as alternative SPE candidates~\cite{sun2024unveiling, chen2026evidence, gavin2026geometry}. Depending on size, structure and symmetry, these complexes are predicted to give rise to shallow defect states close to the conduction band minimum (CBM) in density functional theory (DFT) calculations. The enhanced hybridization between defect state and CBM increases the optical transition dipole and thus activates the defect's optical activity. However, experimental proof for the existence of these states, their proximity to the conduction band of the 2D host, and their responsibility for SPE is so far lacking.

Here we deterministically create sulfur vacancy complexes in monolayer $\mathrm{WS_{2}}$ with the tip of a scanning tunneling microscope (STM) and subsequently manipulate their configuration via tip-induced sulfur diffusion. This serves as a powerful technique to study the effects of system size, geometry and underlying symmetries on electronic in-gap states, as well as their link to optical properties. Various complexes comprising up to four vacancies are characterized using a combination of low-temperature scanning tunneling microscopy (STM), scanning tunneling spectroscopy (STS) and DFT.  In sulfur vacancy dimers, we observe strong hybridization of defect states as a function of vacancy separation and relative orientation to the host lattice. Overall, we experimentally observe the existence of a shallow in-gap state for two vacancy clusters: the lateral nearest-neighbor dimer as well as the vacancy trimer in hollow-centered triangle configuration, thereby confirming theoretical predictions~\cite{sun2024unveiling}.

\section{Results}

\begin{figure*}[]
	\centering
		\includegraphics[width=1.0\textwidth]{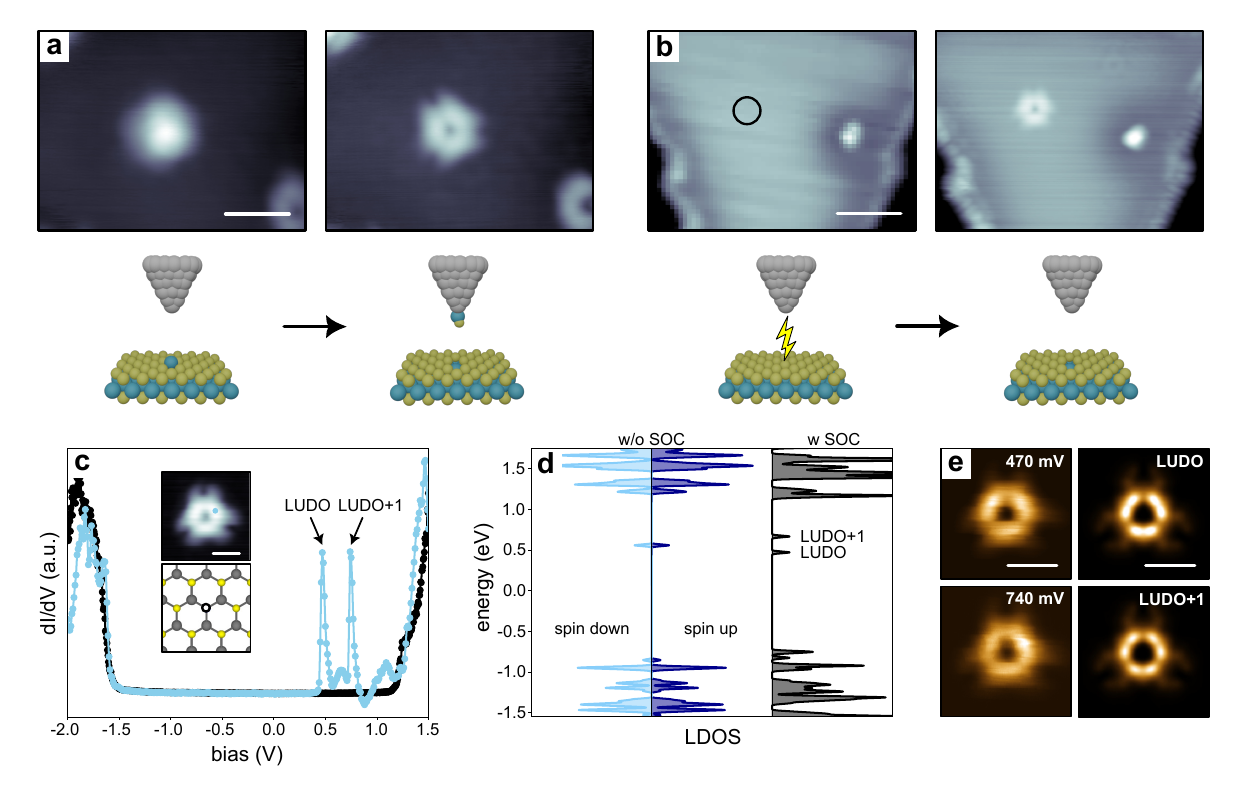}
	\caption{\textbf{Atomic-scale creation of individual sulfur vacancies in monolayer $\mathrm{WS_{2}}$ with the STM tip and their electronic properties.} \textbf{a} STM images and models for the creation of a single sulfur vacancy in $\mathrm{WS_{2}}$ by $\mathrm{W}$ adatom manipulation with the tip (U$_{\text{bias}}= 1.0\, \mathrm{V}$, I$_{\text{set}}=5\, \mathrm{pA}$, scale bar: $2.0\, \mathrm{nm}$). Left: $\mathrm{W}$ atom prior to manipulation. Right: sulfur vacancy after adatom pick-up. \textbf{b} STM images and models for the creation of a sulfur vacancy by a voltage pulse locally applied with the STM tip (U$_{\text{bias}}= 1.3\, \mathrm{V}$, I$_{\text{set}}=40\, \mathrm{pA}$, scale bar: $3.0\, \mathrm{nm}$). Left: pristine $\mathrm{WS_{2}}$ island without defect. The circle marks the voltage pulse position. Right: single sulfur vacancy after the voltage pulse. \textbf{c} Differential conductance point spectra (U$_{\text{bias}}=1.5\, \mathrm{V}$, I$_{\text{set}}=1\, \mathrm{nA}$, U$^{1}_{\text{mod}}=5\, \mathrm{mV}$, U$^{2}_{\text{mod}}=10\, \mathrm{mV}$) measured on pristine $\mathrm{WS_{2}}$ (black) and the sulfur vacancy (blue). The position where the spectrum was measured is indicated in the STM inset (U$_{\text{bias}}=800\, \mathrm{mV}$, I$_{\text{set}}=40\, \mathrm{pA}$, scale bar: $1.0\, \mathrm{nm}$). A ball model is shown below with the sulfur vacancy site marked by the open black circle. \textbf{d} Spin-polarized LDOS without SOC (left) and total LDOS with SOC (right) calculated by DFT. The Fermi level was adjusted to match the experimental LUDO energy. \textbf{e} Comparison of differential conductance maps measured in constant height mode at bias voltages shown in the figure (U$_{\text{mod}}=20\, \mathrm{mV}$, scale bar: $1.0\, \mathrm{nm}$) and DFT simulations (scale bar: $1.0\, \mathrm{nm}$) of LUDO and LUDO+1.}
	\label{figure creation}
\end{figure*}

Monolayer $\mathrm{WS_{2}}$ is grown on a graphene/$\mathrm{Ir}$(111) substrate using molecular-beam epitaxy (Methods and Supplementary Fig.~1). Creation of vacancies is performed locally using the STM tip, either via voltages pulses~\cite{hosoki1992surface,willke2026atomic}, or via vertical manipulation of single $\mathrm{W}$ adatoms deposited at low temperatures, see Fig.~\ref{figure creation}\textbf{a-b}. The latter method has been previously reported in Ref.~\cite{jansen2024tip} for $\mathrm{Fe}$ adatoms on $\mathrm{MoS_{2}}$. On pristine $\mathrm{WS_{2}}$ it has a much higher success rate compared to voltage pulses.

When a chalcogen atom is removed from one of the anion planes of $\mathrm{WS_{2}}$, the surrounding transition metal $d$-levels re-hybridize and produce defect states. They form according to the irreducible representations of the C$_{3v}$ point group, allowing for non-degenerate $a_{1}$, $a_{2}$ as well as doubly-degenerate $e$ states. The latter splits due to large spin-orbit coupling (SOC) and forms the lowest unoccupied defect orbital (LUDO) and a higher energy defect orbital (LUDO+1) deep in the band gap of $\mathrm{WS_{2}}$~\cite{li2016strong, khan2017vacancy, tan2020stability}.

In Fig.~\ref{figure creation}\textbf{c} we present a characteristic differential conductance (d$I$/d$V$) point spectrum acquired on a single top-layer sulfur vacancy, together with a spectrum measured on pristine $\mathrm{WS_{2}}$. The d$I$/d$V$ signal is proportional to the local density of states (LDOS) and thus detects the energetic position of the valence band, conduction band and defect states. In the band gap region of the pristine $\mathrm{WS_{2}}$ sample (black spectrum) we observe two sharp peaks when measuring on a vacancy lobe (blue spectrum). We attribute these peaks to the SOC-split LUDO and LUDO+1 states~\cite{schuler2019large}. The satellite peaks visible at bias voltages above the two states are interpreted as vibronic excitations induced by inelastic tunneling~\cite{schuler2019large, cochrane2021spin}. Our DFT calculations (Methods) confirm the presence of two in-gap states for single sulfur vacancies as shown in Fig.~\ref{figure creation}\textbf{d}, which form spin-degenerate Kramer's pairs. Differential conductance maps recorded at the energetic positions of the LUDO and LUDO+1, presented in Fig.~\ref{figure creation}\textbf{e}, show localized orbitals with 3-fold symmetry. These orbitals are the building blocks for the vacancy complexes discussed in the following. Additional measurements of single sulfur vacancies are shown in Supplementary Figs.~2 and 3.

\begin{figure*}[]
	\centering
		\includegraphics[width=1.0\textwidth]{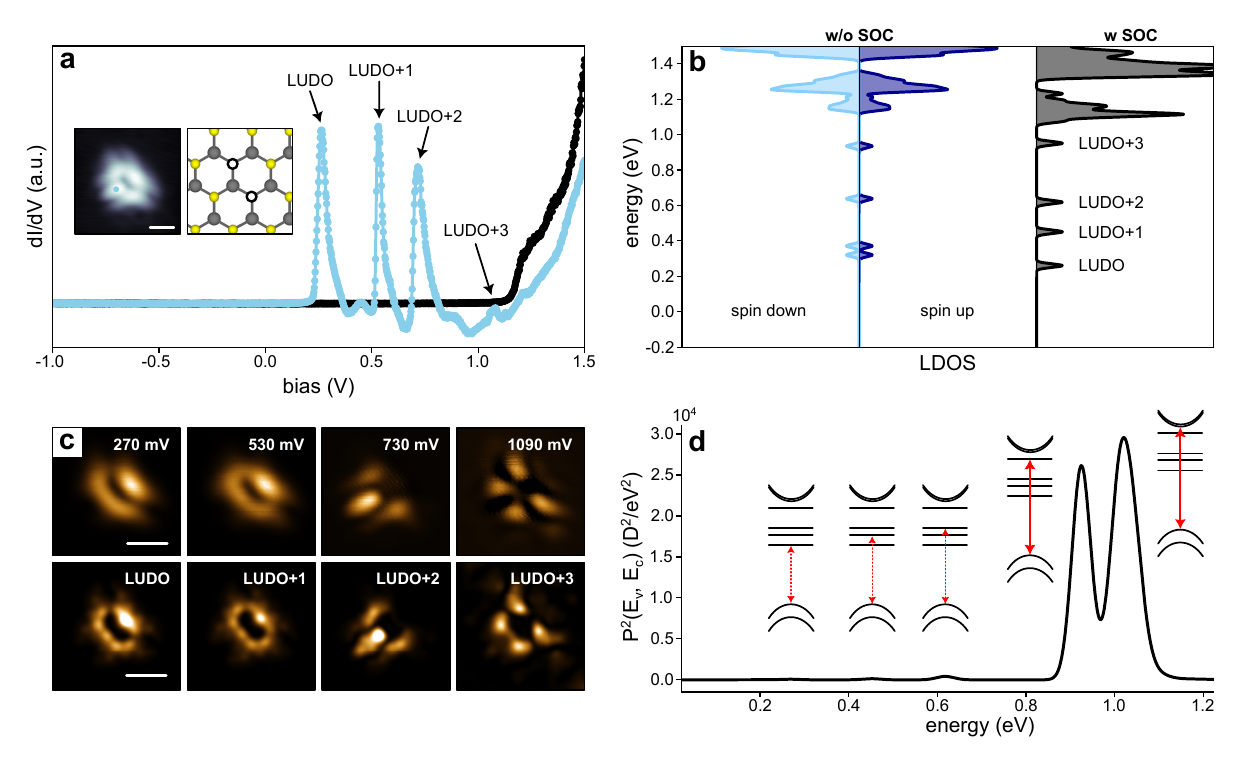}
	\caption{\textbf{Electronic properties of a sulfur vacancy dimer in nearest-neighbor configuration.} \textbf{a} Differential conductance point spectra (U$_{\text{bias}}=1.5\, \mathrm{V}$, I$_{\text{set}}=100\, \mathrm{pA}$, U$_{\text{mod}}=5\, \mathrm{mV}$) measured on the lobes of the dimer (blue) as marked in the STM inset (U$_{\text{bias}}=1.0\, \mathrm{V}$, I$_{\text{set}}=5\, \mathrm{pA}$, scale bar: $1.0\, \mathrm{nm}$) and on pristine $\mathrm{WS_{2}}$ (black). Positions of in-gap states are indicated by arrows. \textbf{b} Computed LDOS of a sulfur vacancy dimer in nearest-neighbor configuration obtained by DFT. On the left, the spin-polarized LDOS excluding SOC and on the right, the total LDOS including SOC is presented. \textbf{c} Comparison of differential conductance maps measured in constant height mode at bias voltages shown in the figure (U$_{\text{mod}}=20\, \mathrm{mV}$, scale bar: $1.0\, \mathrm{nm}$) and DFT simulations (scale bar: $1.0\, \mathrm{nm}$) of the four in-gap states of the dimer. \textbf{d} Optical transition dipoles calculated for transitions between valence band maximum and unoccupied states, consisting of defect orbitals and conduction band. The sketches next to the peaks show valence and conduction band extrema as well as defect states at the K-point of $\mathrm{WS_{2}}$. Optical transitions are indicated by red arrows.}
	\label{figure 1a dimer}
\end{figure*}

Having introduced the properties of an isolated vacancy, we move on to the vacancy dimer in nearest-neighbor configuration. The inset of Fig.~\ref{figure 1a dimer}\textbf{a} shows an STM image and a ball model of the nearest-neighbor dimer. The dimer lowers the point group symmetry from $C_{3v}$ to $C_{s}$, leaving only one mirror symmetry axis that runs vertically to the line connecting the two vacancies. Fig.~\ref{figure 1a dimer}\textbf{a} also shows differential conductance point spectra measured on pristine $\mathrm{WS_{2}}$ (black) and on a dimer lobe (blue). The dimer spectrum shows three pronounced peaks deep in the band gap, together with a weak peak close to the onset of the CBM (see also Supplementary Fig.~4). These peaks are interpreted as in-gap defect states labeled LUDO, LUDO+1, LUDO+2 and LUDO+3. This assignment is corroborated by our DFT calculations in Fig.~\ref{figure 1a dimer}\textbf{b}. Hybridization between the defect orbitals lifts all degeneracies, as  the $C_{s}$ point group, in contrast to the $C_{3v}$ point group of the single vacancy, does not have an $E$ irreducible representation. When considering the effects of SOC, the energy position of all four LUDOs in the computed total LDOS agrees well with our spectroscopic findings. This also holds for the spatial distribution of the four in-gap states, presented in Fig.~\ref{figure 1a dimer}\textbf{c}. The agreement between experimental maps and simulations provides further evidence for the assignment to this defect configuration.

The shallow in-gap state close to the CBM (LUDO+3) has profound impact on the optical properties of the nearest-neighbor sulfur vacancy dimer. To illustrate this, we performed transition dipole calculations for optical transitions throughout the Brillouin zone, see Supplementary Note~1 and Supplementary Fig.~5. The most relevant optical transitions are characterized by the transition dipole moments between the valence band maximum (VBM) and the unoccupied in-gap defect states, as well as between the VBM and the conduction band minimum (CBM), which serves as a reference for the host band-edge transition. They are presented in Fig.~\ref{figure 1a dimer}\textbf{d}. The three peaks with small transition dipoles correspond to transitions between VBM and deep in-gap defect states (LUDO to LUDO+2). They are considered dark, as their intensities are much smaller than the direct band-to-band transitions in pristine $\mathrm{WS_{2}}$ (Supplementary Fig.~6). The shallow state LUDO+3, on the other hand, gives rise to a transition dipole with comparable intensity to the direct band-to-band transition. Our calculations therefore confirm the enhanced optical activity of the nearest-neighbor dimer, arising from hybridization of LUDO+3 with the CBM and consequent capture of transition dipole. Introducing strain shifts the relative energies of CBM and LUDO+3 and simultaneously reduces its transition dipole, confirming the small energy separation of CBM and defect state as the decisive factor for optical activity, see Supplementary Fig.~7.

\begin{figure*}[]
	\centering
		\includegraphics[width=1.0\textwidth]{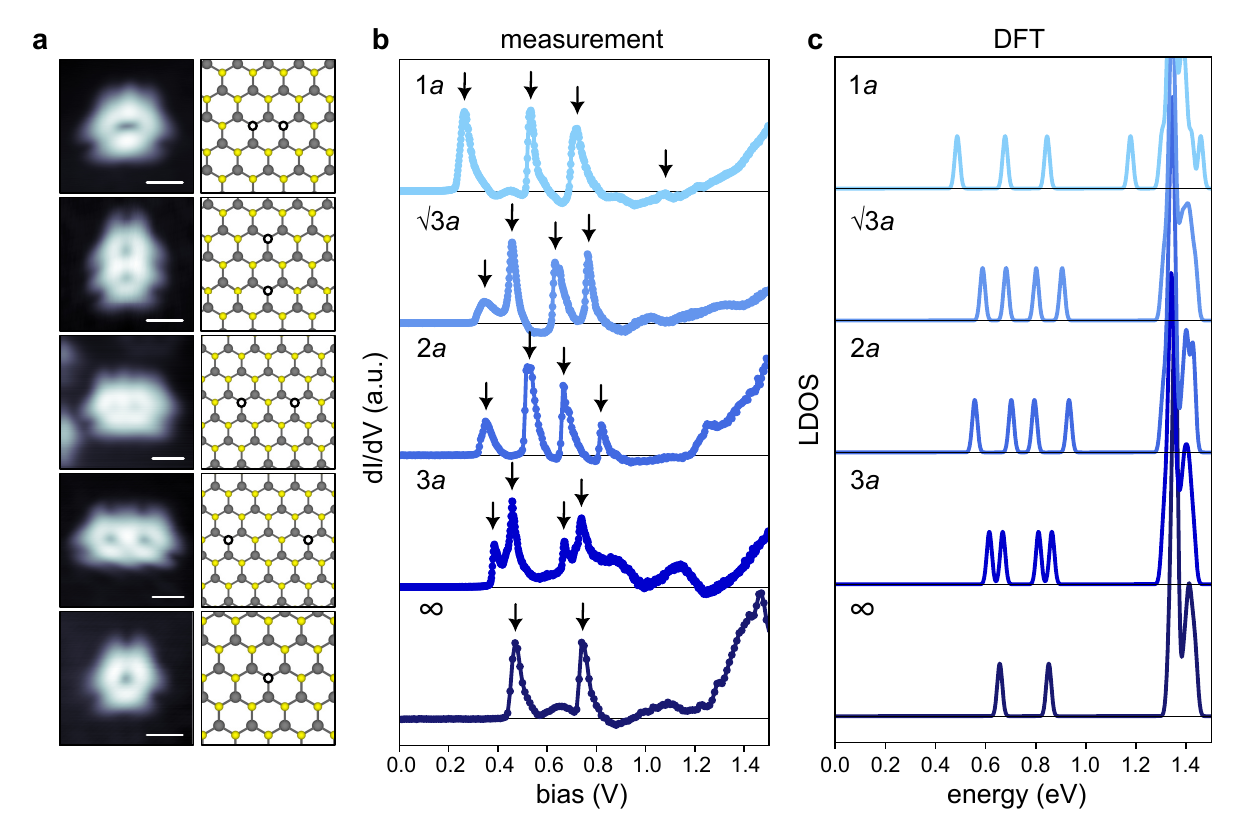}
	\caption{\textbf{Hybridization and splitting of electronic in-gap states in sulfur vacancy dimers as a function of dimer separation in units of the lattice parameter $a$.} \textbf{a} STM images and ball models of sulfur vacancy dimers with different separations that are increasing from top to bottom ($1a$: U$_{\text{bias}}=1.0\, \mathrm{V}$, I$_{\text{set}}=5\, \mathrm{pA}$; $\sqrt{3}a$: U$_{\text{bias}}=1.0\, \mathrm{V}$, I$_{\text{set}}=10\, \mathrm{pA}$; $2a$: U$_{\text{bias}}=600\, \mathrm{mV}$, I$_{\text{set}}=40\, \mathrm{pA}$; $3a$: U$_{\text{bias}}=1.0\, \mathrm{V}$, I$_{\text{set}}=10\, \mathrm{pA}$; $\infty$: U$_{\text{bias}}=1.0\, \mathrm{V}$, I$_{\text{set}}=10\, \mathrm{pA}$, scale bars: $1.0\, \mathrm{nm}$) \textbf{b} Sequence of differential conductance spectra measured on sulfur vacancy dimers ($1a$: U$_{\text{bias}}=1.5\, \mathrm{V}$, I$_{\text{set}}=100\, \mathrm{pA}$, U$_{\text{mod}}=5\, \mathrm{mV}$; $\sqrt{3}a$: U$_{\text{bias}}=1.5\, \mathrm{V}$, I$_{\text{set}}=1\, \mathrm{nA}$, U$_{\text{mod}}=5\, \mathrm{mV}$; $2a$: U$_{\text{bias}}=1.5\, \mathrm{V}$, I$_{\text{set}}=1\, \mathrm{nA}$, U$_{\text{mod}}=5\, \mathrm{mV}$; $3a$: U$_{\text{bias}}=1.5\, \mathrm{V}$, I$_{\text{set}}=1\, \mathrm{nA}$, U$_{\text{mod}}=5\, \mathrm{mV}$; $\infty$: U$_{\text{bias}}=1.5\, \mathrm{V}$, I$_{\text{set}}=1\, \mathrm{nA}$, U$_{\text{mod}}=10\, \mathrm{mV}$). State positions are indicated by arrows for better visibility. \textbf{c} Theoretical LDOS of the vacancy dimers computed by DFT.}
	\label{figure dimers}
\end{figure*}

To uncover the impact of symmetry and separation on the dimer in-gap level structure and its evolution, we created sulfur vacancy dimers with different spacings and orientations, see Fig.~\ref{figure dimers}\textbf{a}. The separation is specified relative to the lattice constant $a$ of $\mathrm{WS_{2}}$. In Fig.~\ref{figure dimers}\textbf{b} differential conductance spectra measured on the lobes of the corresponding dimers are shown in order of increasing separation from top to bottom. Fig.~\ref{figure dimers}\textbf{c} displays the corresponding theoretical LDOS. Spatial maps of all orbitals and additional data can be found in Supplementary Figs.~8 to 10.

The results displayed in Fig.~\ref{figure dimers} provide insights into the hybridization of sulfur vacancies and the positioning of hybrid states in the $\mathrm{WS_{2}}$ band gap. All of the dimers have in common that the four in-gap states are non-degenerate, as $C_{s}$ does not allow for symmetry-caused degeneracies. As the dimer separation decreases, the separation between the defect states increases, as expected for molecular bonding and anti-bonding states. The $\sqrt{3}a$-dimer, however, does not follow this trend; its state separation is smaller than in the $2a$ case. Hence, both separation and orientation of the dimer dictate the magnitude of orbital overlap and consequently its in-gap state separation. Only nearest neighbor dimers have a shallow in-gap state and are thus expected to be optically active.

\begin{figure*}[]
	\centering
		\includegraphics[width=1.0\textwidth]{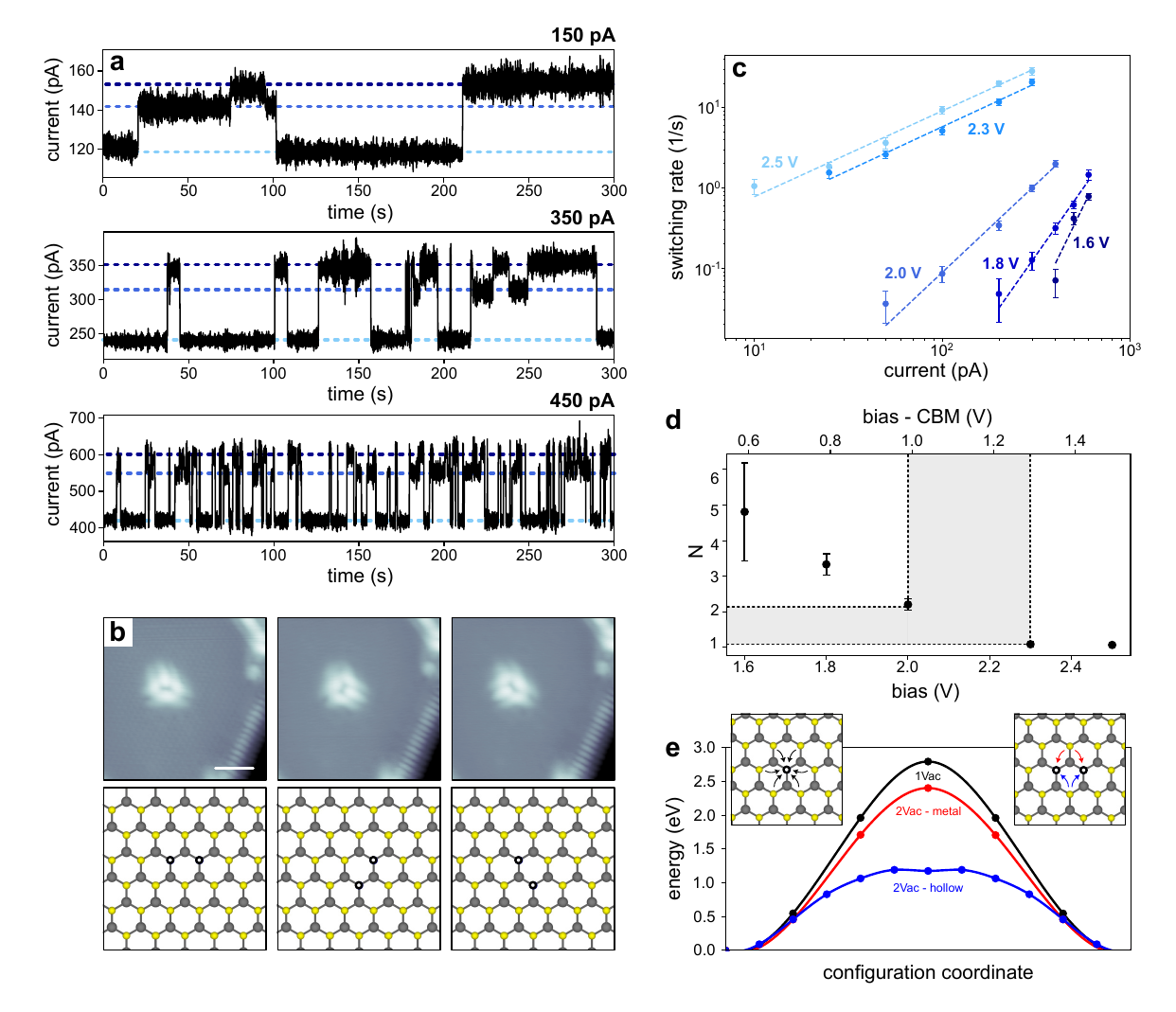}
	\caption{\textbf{Telegraph noise analysis and sulfur diffusion process in nearest-neighbor sulfur vacancy dimers.} \textbf{a} Telegraph noise signals measured on a defect lobe with disabled feedback at $1.8\, \mathrm{V}$ showing three distinct levels indicated by dotted lines. The state switching rates shows dependency on the tunneling current (upper panel: $150\, \mathrm{pA}$, middle panel: $350\, \mathrm{pA}$, lower panel: $450\, \mathrm{pA}$). \textbf{b} STM images (U$_{\text{bias}}=1.0\, \mathrm{V}$, I$_{\text{set}}=5\, \mathrm{pA}$, scale bar: $2.0\, \mathrm{nm}$) and ball models of the vacancy dimer and its immediate environment on the $\mathrm{WS_{2}}$ island in three orientational states. The three states correspond to a rotation of the vacancy dimer within a hollow-centered sulfur triangle spanned by three atomic positions. The sulfur vacancy positions are indicated by open black circles in the models. \textbf{c} Results of the telegraph noise analysis conducted at different bias voltages and tunneling currents (extracted from data with active feedback loop). The double logarithmic plot shows the switching rate $R$ between the orientational states of the dimer as a function of the tunneling current $I$. The data follows the relation $R = I^{N}$, where $N$ is the slope of the fitted linear functions. \textbf{d} Value of $N$ plotted as a function of applied bias voltage (lower x-axis) and maximum effective energy transferred to the lattice by inelastic electrons (upper x-axis). The marked area indicates the range between two data points lying in the $1 \le N \ge 2$ interval. \textbf{e} Energy barriers for sulfur diffusion computed by DFT. The black curve represents diffusion of a neighboring sulfur atom in presence of a single top-layer sulfur vacancy. The red (path over metal atom, 2Vac-metal) and blue (path over hollow site, 2Vac-hollow) curves belong to diffusion processes of sulfur atoms in presence of a nearest-neighbor vacancy dimer. Diffusion paths are indicated in the ball models in respective color.}
	\label{figure sulfur diffusion}
\end{figure*}

While the monomers and dimers are created in a static environment, local introduction of lattice dynamics would enable construction and rearrangement of larger vacancy complexes. Such dynamics are induced in nearest-neighbor dimers when a sufficiently large bias voltage is applied. Fig.~\ref{figure sulfur diffusion}\textbf{a} shows current traces recorded at three fixed tip-sample separations, with the STM tip positioned on the dimer lobes. One can clearly identify switches between three distinct plateaus in the tunneling current, reminiscent of multi-level random telegraph noise. Each switch corresponds to a rotational motion of the dimer within a triangle of three sulfur lattice sites, see Fig.~\ref{figure sulfur diffusion}\textbf{b} and Supplementary Fig.~11.

The frequency of the switches gradually increases as a function of tunneling current. In contrast, switches can be completely turned off when reducing the bias voltage. This points to an inelastic tunneling process, which was investigated by Stipe et al. for the rotation of adsorbed molecules on metal surfaces~\cite{stipe1998inducing, stipe1998coupling}. To quantify the dynamics observed for vacancy dimers in $\mathrm{WS_{2}}$, we analyze the rotation dynamics from telegraph noise~\cite{stipe1998inducing}. We recorded telegraph noise for a range of different bias voltages and tunneling current setpoints on the dimer lobe. From this data we extract the switching rate $R$, see Supplementary Note~2 and Supplementary Fig.~12 for more details. The result is plotted in Fig.~\ref{figure sulfur diffusion}\textbf{c}. Fitting the slopes $N$ for each bias voltage gives us a relation between switching rate and tunneling current, $R=I^N$. Determining the highest bias voltage for which $N\geq2$ enables us to extract the energy barrier $E_{\rm{rot}}$ of the rotational motion, as it marks a regime above which switching is mainly dominated by single electron processes ($N=1$). The range is indicated by the gray shaded area in Fig.~\ref{figure sulfur diffusion}\textbf{d}. As electrons mainly remain in the $\mathrm{WS_{2}}$ conduction band after the inelastic tunneling process, we subtract the CBM (Supplementary Fig.~13) and obtain $E_{\rm{rot}}=(1.14\pm0.15)\, \mathrm{eV}$.
 
DFT sheds light on the origin of the dimer rotation observed in experiment. Fig.~\ref{figure sulfur diffusion}\textbf{e} shows the energy barrier for sulfur diffusion into a neighboring vacancy site. Two inequivalent paths exist for nearest-neighbor vacancy dimers~\cite{komsa2013defects}. The energy barrier via the hollow site (2Vac-hollow, blue) is strongly reduced to $E_{\rm{rot}}=1.19\, \mathrm{eV}$, in good agreement with the experimental value and previous calculations~\cite{wang2024comparative}. In contrast, migration via a $\mathrm{W}$ atom (2Vac-metal, red) has a much higher energy barrier~\cite{kieczka2025dynamics}, close to the value for a single vacancy (black). In consequence, the dynamics of vacancy dimers is bound to three configurations, explaining the tri-stability observed in the telegraph noise. More details on the energy barrier calculations are provided in Supplementary Note~3 and Supplementary Fig.~14.

\begin{figure*}[]
	\centering
		\includegraphics[width=1.0\textwidth]{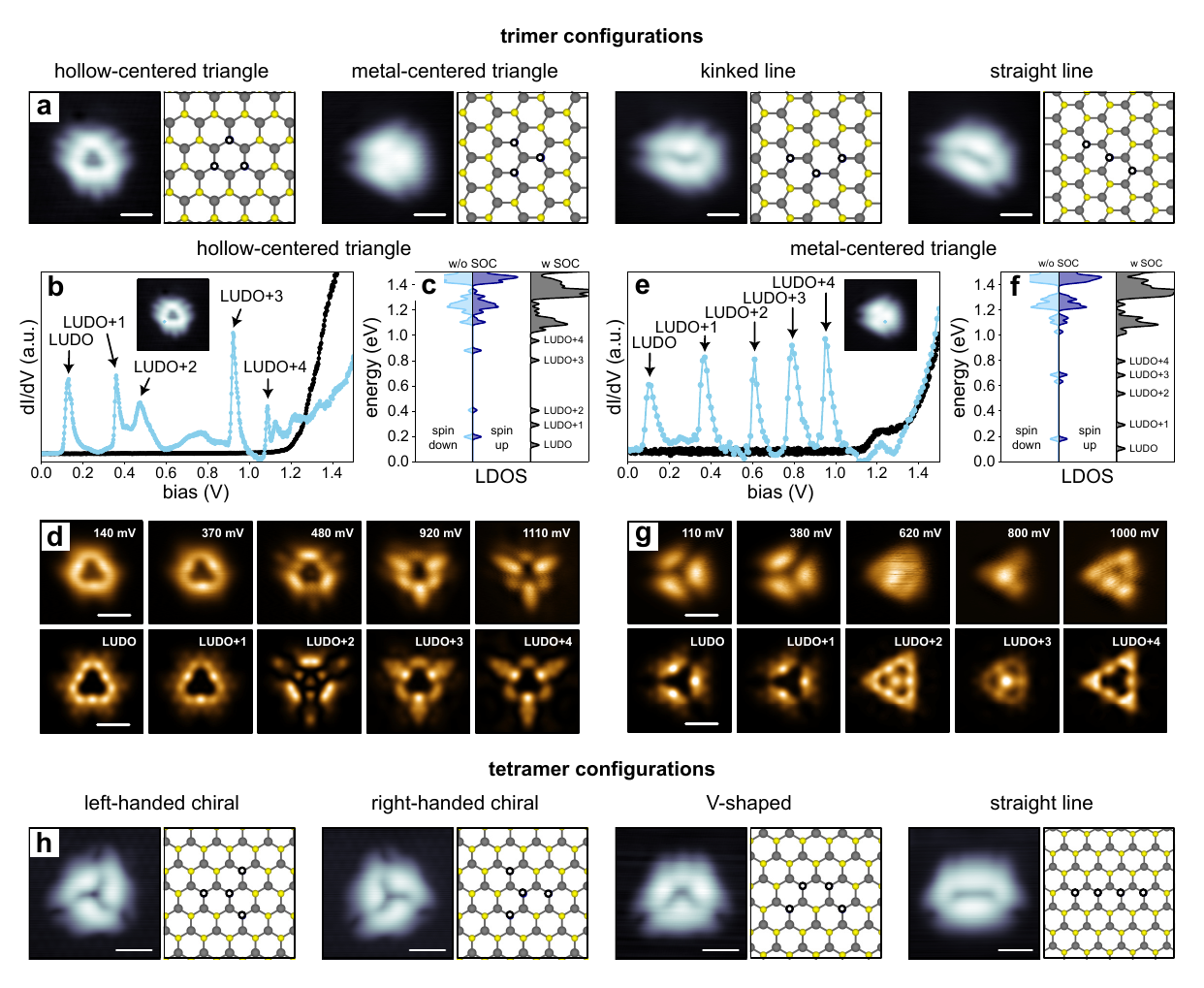}
	\caption{\textbf{Tip-assembled sulfur vacancy trimers and tetramers in different configurations.} \textbf{a} STM images (U$^{1}_{\text{bias}}=900\, \mathrm{mV}$, U$^{2,3,4}_{\text{bias}}=1.0\, \mathrm{V}$, I$_{\text{set}}=5\, \mathrm{pA}$, scale bars: $1.0\, \mathrm{nm}$) and ball models of four different trimer structures: hollow-centered triangle, metal-centered triangle, kinked line and straight line configurations. \textbf{b} Differential conductance spectra measured on pristine $\mathrm{WS_{2}}$ (black) and the defect lobes of the hollow-centered vacancy triangle (blue) (U$_{\text{bias}}=1.5\, \mathrm{V}$, I$_{\text{set}}=100\, \mathrm{pA}$, U$_{\text{mod}}=5\, \mathrm{mV}$). The exact measurement position is marked in the STM image in the inset and LUDO state positions are indicated by arrows. \textbf{c} LDOS calculations as obtained by DFT for a hollow-centered vacancy triangle. On the left, the spin-polarized LDOS excluding SOC and on the right the total LDOS including SOC is shown. The Fermi level is adjusted such that the LUDO energy fits the experimental value. \textbf{d} Comparison between differential conductance maps of the hollow-centered vacancy triangle defect states measured in constant-height mode at indicated bias voltages (U$_{\text{mod}}=20\, \mathrm{mV}$, scale bar: $1.0\, \mathrm{nm}$) and simulated maps (scale bar: $1.0\, \mathrm{nm}$) obtained by DFT of the same defect states. \textbf{e} Differential conductance spectra measured on pristine $\mathrm{WS_{2}}$ (black, U$_{\text{bias}}=1.5\, \mathrm{V}$, I$_{\text{set}}=10\, \mathrm{pA}$, U$_{\text{mod}}=5\, \mathrm{mV}$) and the defect lobes of the metal-centered vacancy triangle (blue, U$_{\text{bias}}=1.5\, \mathrm{V}$, I$_{\text{set}}=10\, \mathrm{pA}$, U$_{\text{mod}}=10\, \mathrm{mV}$). The exact measurement position is marked in the STM image in the inset. \textbf{f} LDOS calculations as obtained by DFT for a metal-centered vacancy triangle. On the left, the spin-polarized LDOS excluding SOC and on the right the total LDOS including SOC is shown. The Fermi level is adjusted such that the LUDO energy fits the experimental value. \textbf{g} Comparison between differential conductance maps of the metal-centered vacancy triangle defect states measured in constant-height mode at indicated bias voltages (U$_{\text{mod}}=20\, \mathrm{mV}$, scale bar: $1.0\, \mathrm{nm}$) and simulated maps (scale bar: $1.0\, \mathrm{nm}$) obtained by DFT of the same defect states. \textbf{h} STM images (U$^{1,2}_{\text{bias}}=500\, \mathrm{mV}$, U$^{3,4}_{\text{bias}}=1.0\, \mathrm{V}$, I$_{\text{set}}=40\, \mathrm{pA}$, scale bars: $1.0\, \mathrm{nm}$) and models of four different tetramer structures: left- and right-handed chiral, V-shaped and straight line configurations.}
	\label{figure trimers and tetramers}
\end{figure*}

With the ability to create and subsequently manipulate the position of vacancies in larger complexes, we extend our toolbox of in-gap state engineering further by adding more sulfur vacancies to the complex. The next larger complex is the sulfur vacancy trimer, which consists of three closely spaced vacancies. Once a trimer is created, additional sulfur migration paths with lowered energy barriers enable us to generate various configurations using the same vacancy complex. In total, the three vacancies can now explore a triangular area of $\mathrm{WS_{2}}$ that contains six sulfur sites, as theoretically discussed in Ref.~\cite{flototto2026cooperative}. In Fig.~\ref{figure trimers and tetramers}\textbf{a} we show a collection of four subsequent trimer configurations with corresponding ball models. The two triangular trimers belong to the $C_{3v}$ point group. All other configurations have lower symmetry, i.e., $C_{s}$ (straight line) or $C_{1}$ (kinked line).
Electronically, each trimer configuration has six hybridized states, two contributed by each of the three sulfur vacancies. Five of these orbitals are located within the band gap of $\mathrm{WS_{2}}$. 
In the following we focus on the electronic properties of the two triangular trimers. A full characterization of the line-complexes is provided in Supplementary Figs.~15 and 16. 

Fig.~\ref{figure trimers and tetramers}\textbf{b} shows a differential conductance point spectrum measured on the hollow-centered triangle trimer. The peaks identified as defect states are indicated by arrows and labeled LUDO to LUDO+4. The differential conductance spectrum can be compared to DFT calculations, presented in Fig.~\ref{figure trimers and tetramers}\textbf{c}. Two peaks in the LDOS without SOC can be identified as doubly-degenerate states enforced by the $C_{3v}$ point group symmetry. These degeneracies are lifted when including SOC. In Fig.~\ref{figure trimers and tetramers}\textbf{d} we present differential conductance maps of the five in-gap states of the hollow-centered triangle trimer and compare them to maps simulated by DFT. We find excellent agreement between experiment and theory, justifying the assignment of in-gap states. Interestingly the two pairs of states (LUDO and LUDO+1; LUDO+3 and LUDO+4) that are degenerate in absence of SOC according to our DFT calculations, each have the same orbital appearance, in analogy to LUDO and LUDO+1 of the monomer. In contrast, LUDO+2, which is non-degenerate in absence of SOC, does not have such a twin state. The hollow-centered triangle trimer hosts a shallow state (LUDO+4), which is reminiscent of the situation in the nearest-neighbor dimer and points to optical activity. Our optical transition dipole calculations indeed find strong transitions from LUDO+4 to VBM states (Supplementary Fig.~17), further supporting that defect-CBM hybridization mediates optical activity.

The in-gap state structure of the metal-centered triangle trimer can be inferred from the differential conductance spectrum presented in Fig.~\ref{figure trimers and tetramers}\textbf{e}. In Fig.~\ref{figure trimers and tetramers}\textbf{f} we show DFT calculations for the LDOS without (left) and with SOC (right). Also here we can see that without SOC two pairs of degenerate states exist that split under the influence of SOC. Fig.~\ref{figure trimers and tetramers}\textbf{g} shows a comparison between differential conductance maps of all five in-gap states and their corresponding DFT simulations with a high level of agreement. LUDO and LUDO+1 can be clearly identified as states originating from the same degenerate level by their appearance. Our calculations further reveal that LUDO+2 and LUDO+3 are mixtures of $a$ and $e$ states while LUDO+4 is of pure $e$ state origin.

When adding another sulfur vacancy to the complex, a tetramer is created, which can explore in total 10 lattice sites~\cite{flototto2026cooperative}. In Fig.~\ref{figure trimers and tetramers}\textbf{h} we highlight four tetramer configurations that we obtain from the same complex: two configurations with a chiral vacancy structure in left- and right-handed form, a V-shaped variant and a straight line tetramer. Electronic characterizations of the tetramer in each of the configurations are provided in the Supplementary Figs.~18 to 21. Their energetics are discussed in Supplementary Note~4 and Supplementary Table~1. All tetramers possess states close to the CBM, but in contrast to smaller complexes, these appear less localized due to their strong orbital overlap, marking a transition from localized orbitals to delocalized quasiparticles.

\section{Summary}
In summary, we have shown atomic-scale assembly of point defect complexes in two-dimensional semiconductors as a powerful tool for engineering electronic in-gap states. Using scanning tunneling microscopy and spectroscopy, we have created, measured and manipulated sulfur vacancy complexes in monolayer $\mathrm{WS_{2}}$, going from single sulfur vacancies to dimers, trimers and finally tetramers. Tip-induced sulfur diffusion effectively rearranges sulfur vacancies in the complexes, enabling the exploration of a large set of configurations. The experimental results are corroborated by density functional theory calculations, which shed light on symmetry-related degeneracies, spin-orbit coupling, as well as optical activity. Among the created sulfur vacancy complexes, we find two single photon emitter candidates, the nearest-neighbor dimer and the hollow-centered triangle trimer, for which we provide a detailed characterization. Especially, we precisely measure the decisive requirement for light emission - the energetic position of hybridized in-gap states relative to the conduction band minimum. Our work establishes tip-induced defect patterning as a viable route to engineer custom quantum states in two-dimensional semiconductors. Ultimately, our bottom-up technique could be used to engineer electronic bands via assembly of sulfur vacancies into lattices.

\section{Methods}

\subsection{Sample preparation}
Cleaning of the $\mathrm{Ir}$(111) crystal is done by subsequent cycles of $1.5\, \mathrm{keV}$ Ar-ion beam bombardment, glowing in an oxygen atmosphere of $9.0 \cdot 10^{-9}\, \mathrm{mbar}$ at $1120\, \mathrm{K}$ and vacuum annealing to $1550\, \mathrm{K}$. Afterwards single-crystal, high-quality graphene is grown on the clean $\mathrm{Ir}$(111) surface by chemical vapor deposition as described in Ref.~\cite{coraux2008structural}. $\mathrm{WS_{2}}$ is then grown on the graphene/$\mathrm{Ir}$(111) substrate by molecular beam epitaxy~\cite{hall2018molecular} using an electron-beam evaporator for $\mathrm{W}$ and a Knudsen cell filled with pyrite (FeS$_{2}$) for sulfur supply. During the growth time of $55\, \mathrm{min}$ the sample is held at $570\, \mathrm{K}$ and exposed to a sulfur pressure of $8.0 \cdot 10^{-9}\, \mathrm{mbar}$. After growth the sample is annealed at $1070\, \mathrm{K}$ in $5.0 \cdot 10^{-9}\, \mathrm{mbar}$ sulfur atmosphere for $120\, \mathrm{s}$. To remove excess sulfur at the edges of the islands, the sample is annealed again at slightly lower temperatures of $970\, \mathrm{K}$ for $120\, \mathrm{s}$ in vacuum after the chamber has returned to its base pressure. Evaporation of single $\mathrm{W}$ atoms onto the surface was done in-situ through the radiation shielding of the STM at $7\, \mathrm{K}$ for $120\, \mathrm{s}$ using an electron-beam evaporator.

\subsection{Computational Details}

Density functional theory (DFT) calculations are done in the projector-augmented wave formalism of the VASP software package~\cite{vasp1, vasp2, vasp3}. The Perdew-Burke-Ernzerhof (PBE) functional with van der Waals interactions accounted for by Grimme's D2 corrections and a plane-wave basis defined by $500\, \mathrm{eV}$ cutoff are used~\cite{pbe, grimmed2}. The sulfur vacancy complexes are all modeled in a $10 \times 10$ supercell and sampled at the $\Gamma$-point. Relaxations are done without spin-polarization and afterwards single-shot calculations with spin-orbit coupling (SOC) and $350\, \mathrm{eV}$ cutoff are performed. All simulated images are obtained at a height of about $2.9$~\AA~to mimic realistic conditions of the STM experiment. The simulated images presented in the manuscript and the Supplementary Information are adjusted to have the same real space dimensions as the experimental maps .

In order to properly probe the hybridization between the conduction band minimum (CBM) and the defect states in optical transition dipole moment calculations, the wave vector $k$ of the CBM (K-point) must be included in the sampled $k$-point mesh. The K-point of the primitive cell falls into the K-point of the $10\times10$ supercell used in the calculations, while only the $\Gamma$-point was used in the previous calculations. Thus we determine optical properties via non-self-consistent calculations, similar to the conventional approach used for band structures: we first carry out a self-consistent calculation with SOC and then recalculated the wave function at the K-point with a fixed charge density. The transition dipole map adopts the same procedure but recalculated using a $6\times6$ $k$-point mesh, containing both $\Gamma$- and K-point, with a fixed charge density. The data from all $k$-points are then summed up with the proper $k$-point weights. Note, that only the irreducible wedge of the Brillouin zone from collinear calculations is used. While SOC breaks inversion symmetry, density of states (DOS) and transition dipole can be integrated using only the irreducible wedge.

\section{Acknowledgements}
The authors acknowledge helpful discussions with Annika Kurzmann. The authors thank CSC–IT Center for Science Ltd. for generous grants of computer time.

\section{Funding}
The authors acknowledge funding from Deutsche Forschungsgemeinschaft (DFG, German Research Foundation) through CRC 1238 (project number 277146847, subprojects A01 and B06).

\section{Author contributions}
W.J., H.-P.K. and D.J. conceived the project. D.J., G.M. and J.-H. K. prepared the samples and conducted the experiments. H.-P.K. performed the DFT calculations. D.J., J.-H.K., W.J. and H.-P.K. analyzed and interpreted the data. W.J. and T.M. supervised the project. D.J., W.J. and H.-P.K. wrote the manuscript. All authors participated in discussions and contributed to the manuscript.

\section{Data availability}
The data that support the findings of this study are available in Zenodo.

\section{Competing interests}
The authors declare no competing interests.

\section{References}

\bibliography{./library_merged}

\clearpage

\clearpage                          
\hypersetup{pageanchor=false}       
\onecolumngrid
\pagestyle{plain}  
\setcounter{page}{1}
\thispagestyle{empty}   
\addtocontents{toc}{\protect\setcounter{tocdepth}{3}}

\setcounter{figure}{0}   \renewcommand{\thefigure}{S\arabic{figure}}
\setcounter{table}{0}    \renewcommand{\thetable}{S\arabic{table}}
\setcounter{equation}{0} \renewcommand{\theequation}{S\arabic{equation}}

\captionsetup[figure]{font={normalsize,stretch=1.5}}

\DeclareGraphicsExtensions{.pdf}

\begin{center}
{\large\bfseries Supplementary Information:\\[0.5ex]
Atomic-scale assembly of sulfur vacancy complexes\\
for tailored in-gap states in monolayer WS$_{2}$}\\[3ex]

Daniel Jansen,$^{1,*}$ Guangyao Miao,$^{1}$ Jan-Heinrich Keienburg,$^{1}$ Jeison\\
Fischer,$^{1}$ Thomas Michely,$^{1}$ Hannu-Pekka Komsa,$^{2}$ and Wouter Jolie$^{1,3, \dagger}$\\[2ex]

 {\normalfont\small\itshape
   $^{1}$Institute of Physics II, University of Cologne,\\
   Z\"{u}lpicher Stra\ss e 77, 50937 Cologne, Germany\\
   $^{2}$Faculty of Information Technology and Electrical Engineering,\\
   University of Oulu, Pentti Kaiteran katu 1, 90014 Oulu, Finland\\
   $^{3}$Institute for Integrated Natural Sciences,\\
   Physics Department, University of Koblenz,\\
   Universit\"{a}tsstra\ss e 1, 56070 Koblenz, Germany\\}
\end{center}

\vfill
\noindent\rule{0.1\textwidth}{0.4pt}\\[0.5ex]
{\small $^{*}$jansen@ph2.uni-koeln.de}\\
{\small $^{\dagger}$wjolie@ph2.uni-koeln.de}\\

\clearpage

\begingroup
\pagestyle{plain}
\tableofcontents
\endgroup

\newpage

\suppfig{\textbf{Sample preparation: $\mathrm{WS_{2}}$ on graphene/$\mathrm{Ir}$(111)}}{
\begin{figure*}[h!]
	\centering
		\includegraphics[width=1.0\textwidth]{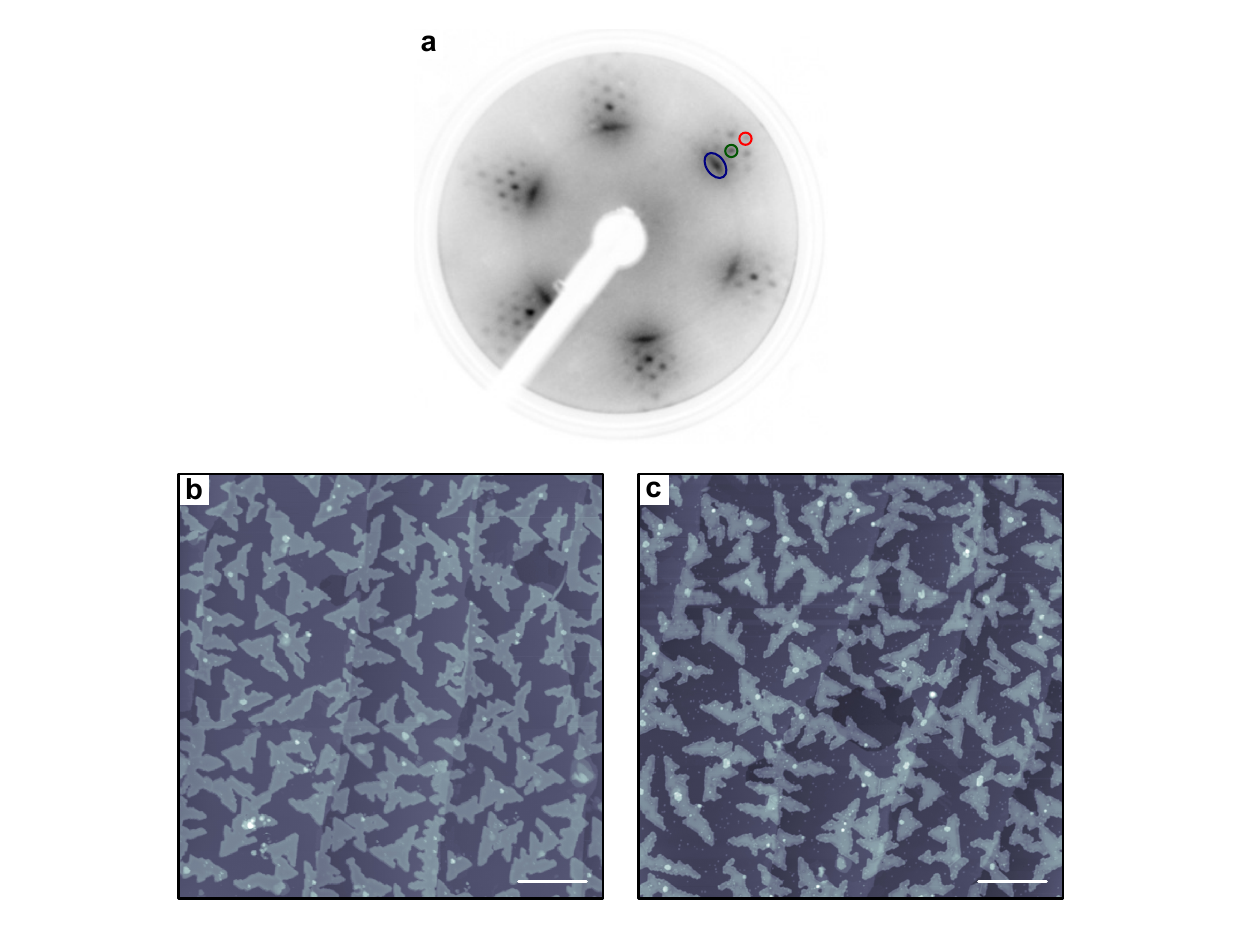}
	\caption{\textbf{Sample preparation: $\mathrm{WS_{2}}$ on graphene/$\mathrm{Ir}$(111).} \textbf{a} Low-energy electron diffraction image of the $\mathrm{WS_{2}}$ on graphene/$\mathrm{Ir}$(111) sample recorded at an electron beam energy of $122.5\, \mathrm{eV}$. Spots belonging to $\mathrm{WS_{2}}$, $\mathrm{Ir}$(111) and graphene are encircled in blue, green and red respectively. The satellite spots are emerging due to the graphene/$\mathrm{Ir}$(111) moir\'e superstructure. \textbf{b} Overview STM image (U$_{\text{bias}}=-2.0\, \mathrm{V}$, I$_{\text{set}}=3\, \mathrm{pA}$, scale bar: $50\, \mathrm{nm}$) of the pristine $\mathrm{WS_{2}}$ on graphene/$\mathrm{Ir}$(111) sample. The total surface coverage is approximately $50\%$, consisting predominantly of monolayers with a small number of bilayers on top.\textbf{c} Overview STM image (U$_{\text{bias}}=1.3\, \mathrm{V}$, I$_{\text{set}}=3\, \mathrm{pA}$, scale bar: $50\, \mathrm{nm}$) of the same sample after evaporation of $\mathrm{W}$ at low temperature.}
	\label{figure sample preparation}
\end{figure*}
}

\clearpage

\suppfig{\textbf{Localization of sulfur vacancy in-gap states}}{
\begin{figure*}[h!]
	\centering
		\includegraphics[width=0.8\textwidth]{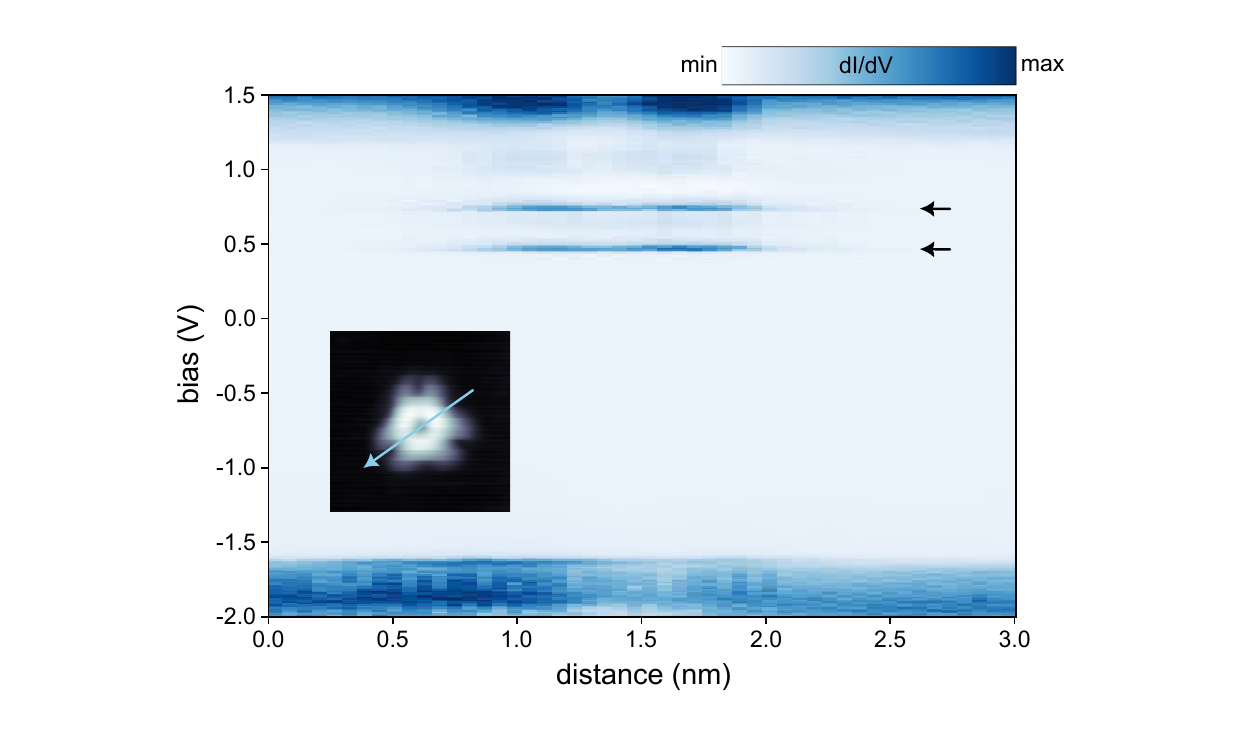}
	\caption{\textbf{Localization of sulfur vacancy in-gap states.} Differential conductance line scan acquired along a line over a sulfur vacancy (U$_{\text{mod}}=10\, \mathrm{mV}$). Arrows mark the energetic positions of in-gap states.}
	\label{figure monomer linescan}
\end{figure*}
}

\clearpage

\suppfig{\textbf{Sulfur vacancy environment on a $\mathrm{WS_{2}}$ island imaged at different bias voltages}}{
\begin{figure*}[h!]
	\centering
		\includegraphics[width=1.0\textwidth]{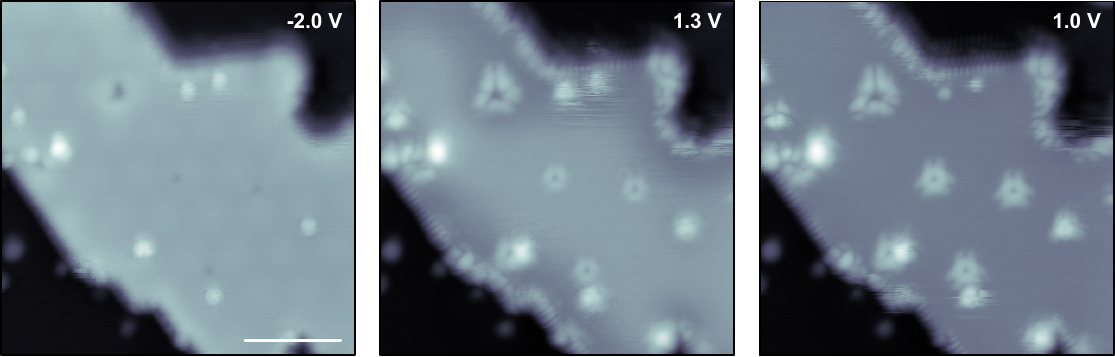}
	\caption{\textbf{Sulfur vacancy environment on a $\mathrm{WS_{2}}$ island imaged at different bias voltages.} Bias voltages as indicated in figure, I$_{\text{set}}=40\, \mathrm{pA}$, scale bar: $5.0\, \mathrm{nm}$. Areas in $\mathrm{WS_{2}}$ affected by band bending close to island edges and defects become visible in the STM image measured at $1.3\, \mathrm{V}$ (center). Sulfur vacancies are created in the defect-free areas which remain unaffected by band bending effects.}
	\label{figure monomer environment}
\end{figure*}
}

\clearpage

\suppfig{\textbf{Localization of nearest-neighbor sulfur vacancy dimer in-gap states}}{
\begin{figure*}[h!]
	\centering
		\includegraphics[width=1.0\textwidth]{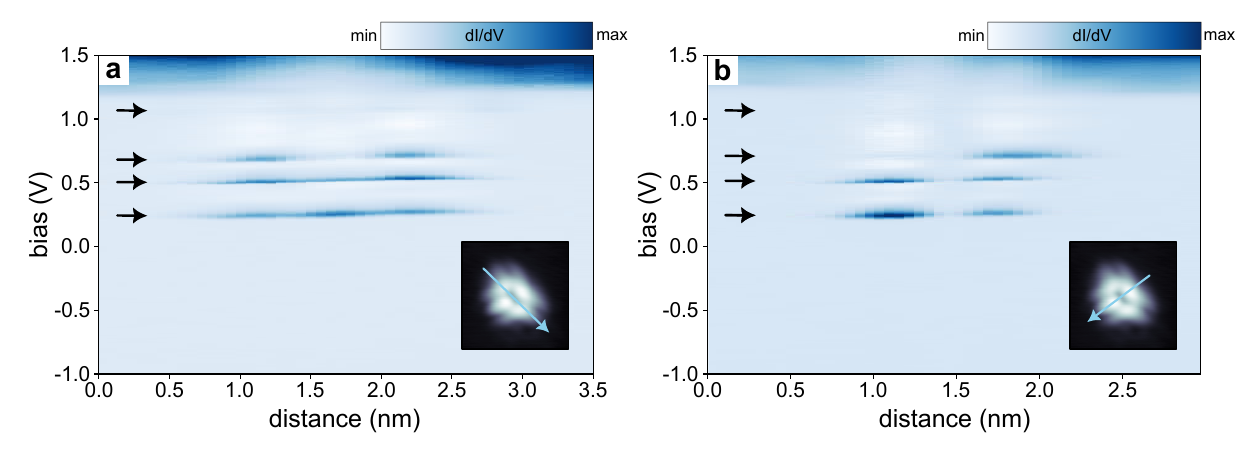}
	\caption{\textbf{Localization of nearest-neighbor sulfur vacancy dimer in-gap states.} Differential conductance line scans (U$_{\text{mod}}=5\, \mathrm{mV}$) measured in constant height mode across the dimer in horizontal \textbf{a} and vertical \textbf{b} directions as indicated by the arrows in the inset STM images (U$_{\text{bias}}=1.0\, \mathrm{V}$, I$_{\text{set}}=5\, \mathrm{pA}$, image size: $4.0 \times 4.0\, \mathrm{nm^{2}}$). Arrows mark the energetic position of in-gap states.}
	\label{figure 1a-dimer linescans}
\end{figure*}
}

\clearpage

\suppnote{\textbf{Optical transition dipole maps}}{
\noindent Optical transition dipole calculations are shown in a color map representation that can be understood in the following way. On the y-axis, the energy range of defect and conduction band (CB) states and on the x-axis the energy range of valence band (VB) states is shown. Next to the color map, DOS plots of the regions are presented for orientation. The color scale in the map gives the squared transition dipole density $\mathrm{P^{2}}(E_{v}, E_{c})$ in units of $\mathrm{D^{2}/eV^{2}}$ (referred to as transition dipole from now on) for transitions from a given defect/CB state to a VB state. An explanatory sketch is shown in Fig.~\ref{figure dipole maps explanation}.\\

\noindent The transition dipole map for pristine monolayer $\mathrm{WS_{2}}$ is presented in Fig.~\ref{figure dipole map bulk}. The dispersive structures arise due to band-to-band transitions at different parts of the Brillouin zone. The two branches stretching towards lower x-axis energies terminate at two spots which correspond to CBM to valence band minimum (VBM) transitions at the K-point. Due to spin-orbit coupling, the CBM is split into two bands, CBM and CBM+1 and the VBM is split into VBM and VBM-1 bands at the K-point. The magnitude of splitting is different for CBM and VBM (approx. $30\, \mathrm{meV}$ and $430\, \mathrm{meV}$~\citeSM{koperski2017optical_merged}) and can be seen in the map as offsets on the axes (VBM splitting on x-axis, CBM splitting on y-axis). The single bright spot seen on the lower right corresponds to transitions at the Q-point.\\

\noindent Fig.~\ref{figure dipole maps nearest neighbor strain} shows the results of the calculations for strain in presence of a nearest-neighbor sulfur vacancy dimer. In \textbf{a} the unstrained case is shown, where the bright high-energy defect state LUDO+3 can be seen just below the CBM. The low-energy defect states LUDO to LUDO+2 give rise to horizontal stripes in the map that can be interpreted as transitions across the Brillouin zone. The magnitude of these transition dipoles is much lower compared to the band-to-band transitions and can therefore be considered dark. Fig.~\ref{figure dipole maps nearest neighbor strain}\textbf{b} presents the results for $2\%$ of tensile strain. The K-point moves lower in energy and the separation to the Q-point increases. As can be seen, tensile strain strongly influences the position of LUDO+3 with respect to the CB states, shifting it well inside the CB. In the map a smearing can be seen at the intersection with the defect state, however a clear distinction between influences of CB and defect states is not possible, due to strong hybridization. When $2\%$ of compressive strain are applied, the K-point moves above the Q-point and LUDO+3 hybridizes with the Q-point states, giving them large transition dipoles.\\

\noindent For the hollow-centered triangle sulfur vacancy trimer, the results of the optical transition dipole calculations are presented in Fig.~\ref{figure dipole map VS3(tri-hole)}. The low-energy defect states LUDO to LUDO+3 give rise to horizontal stripes in the map. The magnitude of the transition dipoles is much lower compared to the band-to-band transitions and are therefore considered dark. The highest energy defect state LUDO+4 on the other hand is situated very close to the CBM and gives rise to bright spots in the map at the position of CBM-to-VBM transitions and is therefore identified as an optically active, bright defect state.
}

\clearpage

\suppfig{\textbf{Explanatory sketch for optical transition dipole maps}}{
\begin{figure*}[h!]
	\centering
		\includegraphics[width=1.0\textwidth]{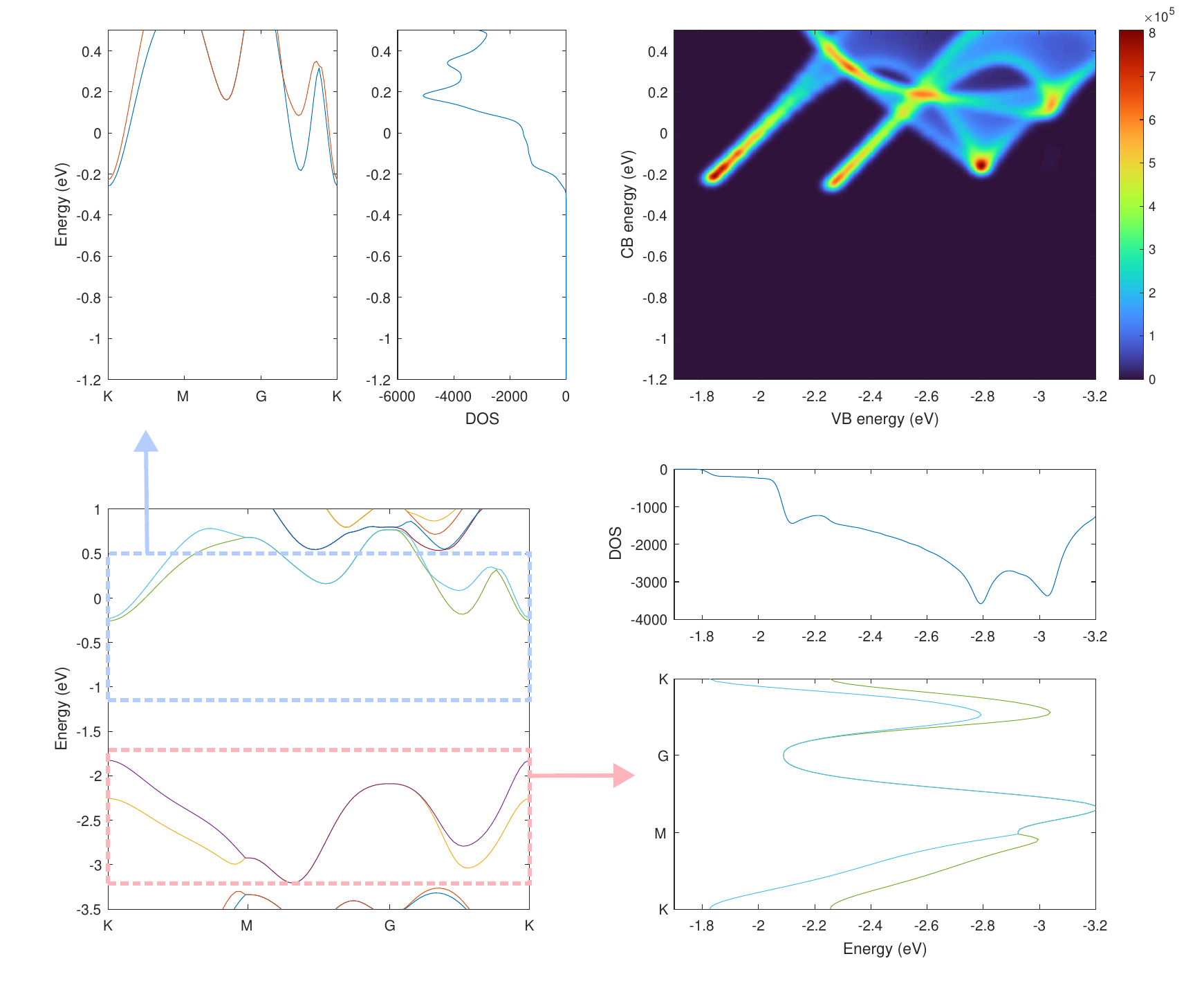}
	\caption{\textbf{Explanatory sketch for optical transition dipole maps.} \textbf{Bottom left:} Band structure plot of monolayer $\mathrm{WS_{2}}$ along K-M-$\Gamma$-K path. \textbf{Top left:} Conduction band energy region of the band structure with corresponding density of states plot next to it. \textbf{Bottom right:} Valence band energy region of the band structure with corresponding density of states plot next to it. \textbf{Top right:} Color map for optical dipole moments (color scale) for transitions between conduction band states (y-axis) and valence band states (x-axis).}
	\label{figure dipole maps explanation}
\end{figure*}
}

\clearpage

\suppfig{\textbf{Optical transition dipole map of pristine $\mathrm{WS_{2}}$.}}{
\begin{figure*}[h!]
	\centering
		\includegraphics[width=0.8\textwidth]{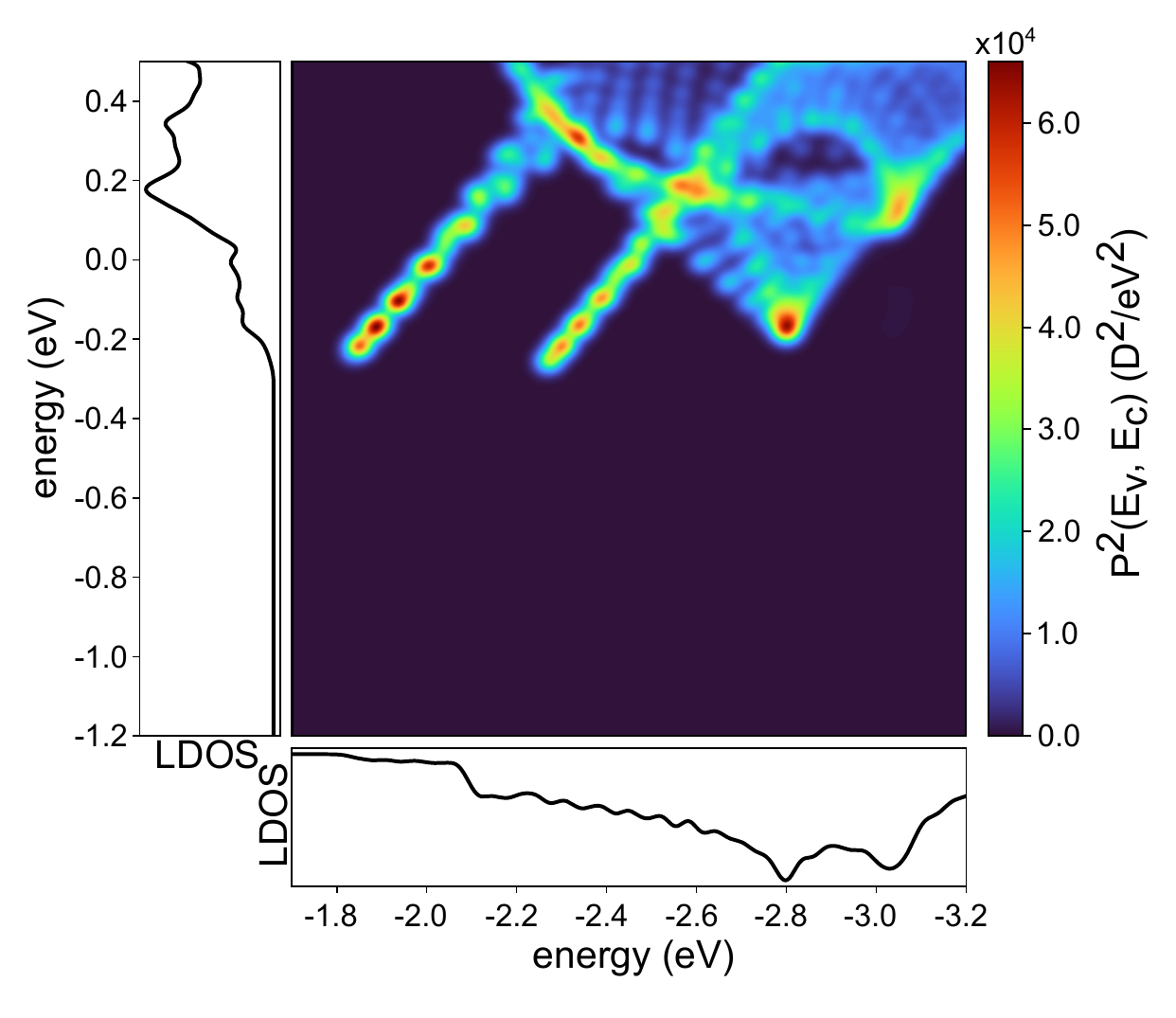}
	\caption{\textbf{Optical transition dipole map of pristine $\mathrm{WS_{2}}$.}}
	\label{figure dipole map bulk}
\end{figure*}
}

\clearpage

\suppfig{\textbf{Influence of strain on the optical transition dipole of the nearest-neighbor sulfur vacancy dimer}}{
\begin{figure*}[h!]
	\centering
		\includegraphics[width=1.0\textwidth]{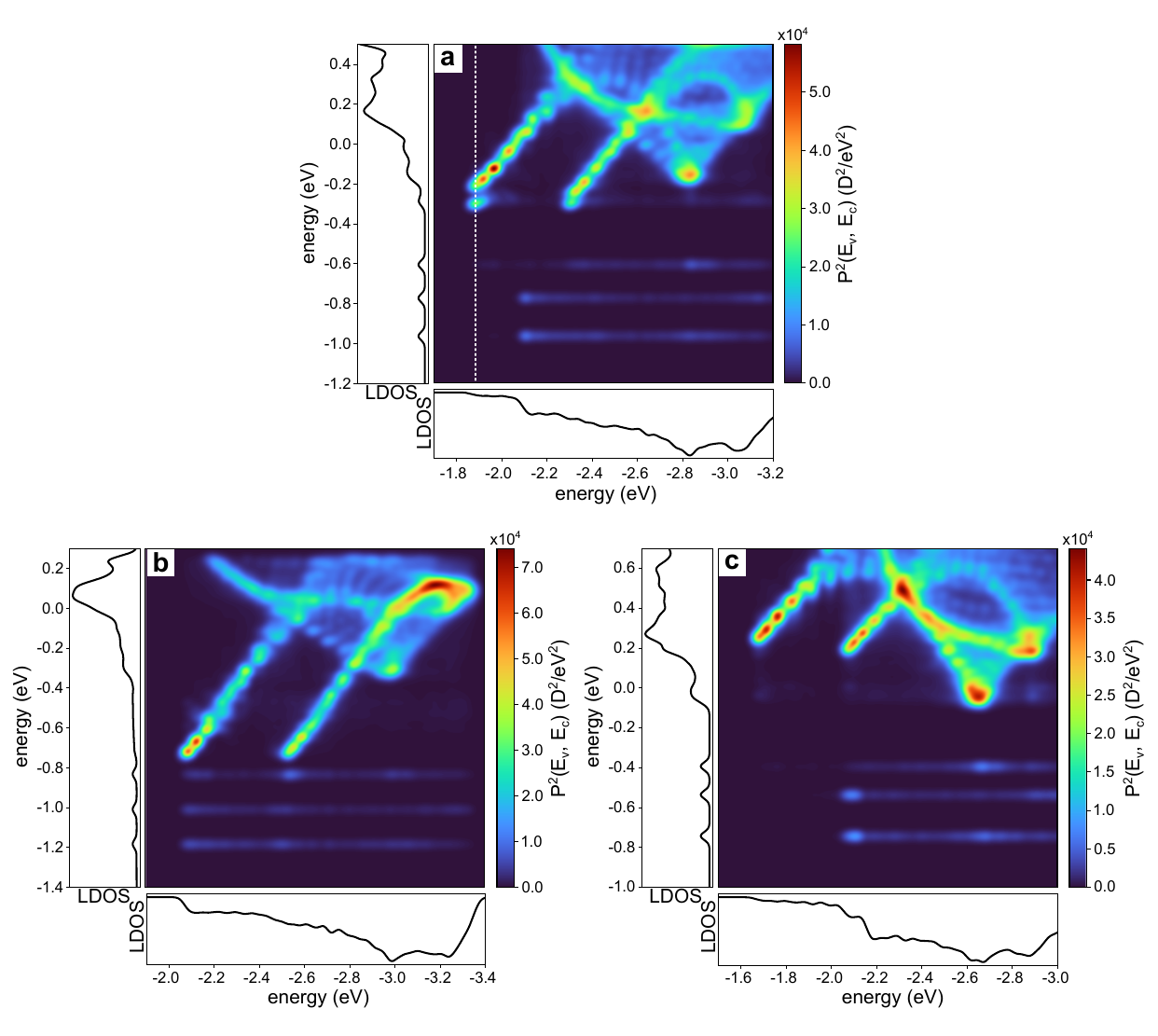}
	\caption{\textbf{Influence of strain on the optical transition dipole of the nearest-neighbor sulfur vacancy dimer.} \textbf{a} Unstrained case. The white dashed line indicates the transition dipoles discussed in the main text. \textbf{b} Case for $2\%$ tensile strain. \textbf{c} Case for $2\%$ compressive strain.}
	\label{figure dipole maps nearest neighbor strain}
\end{figure*}
}

\clearpage

\suppfig{\textbf{Electronic characterization of a sulfur vacancy dimer in a $\sqrt{3}a$-configuration}}{
\begin{figure*}[h!]
	\centering
		\includegraphics[width=0.8\textwidth]{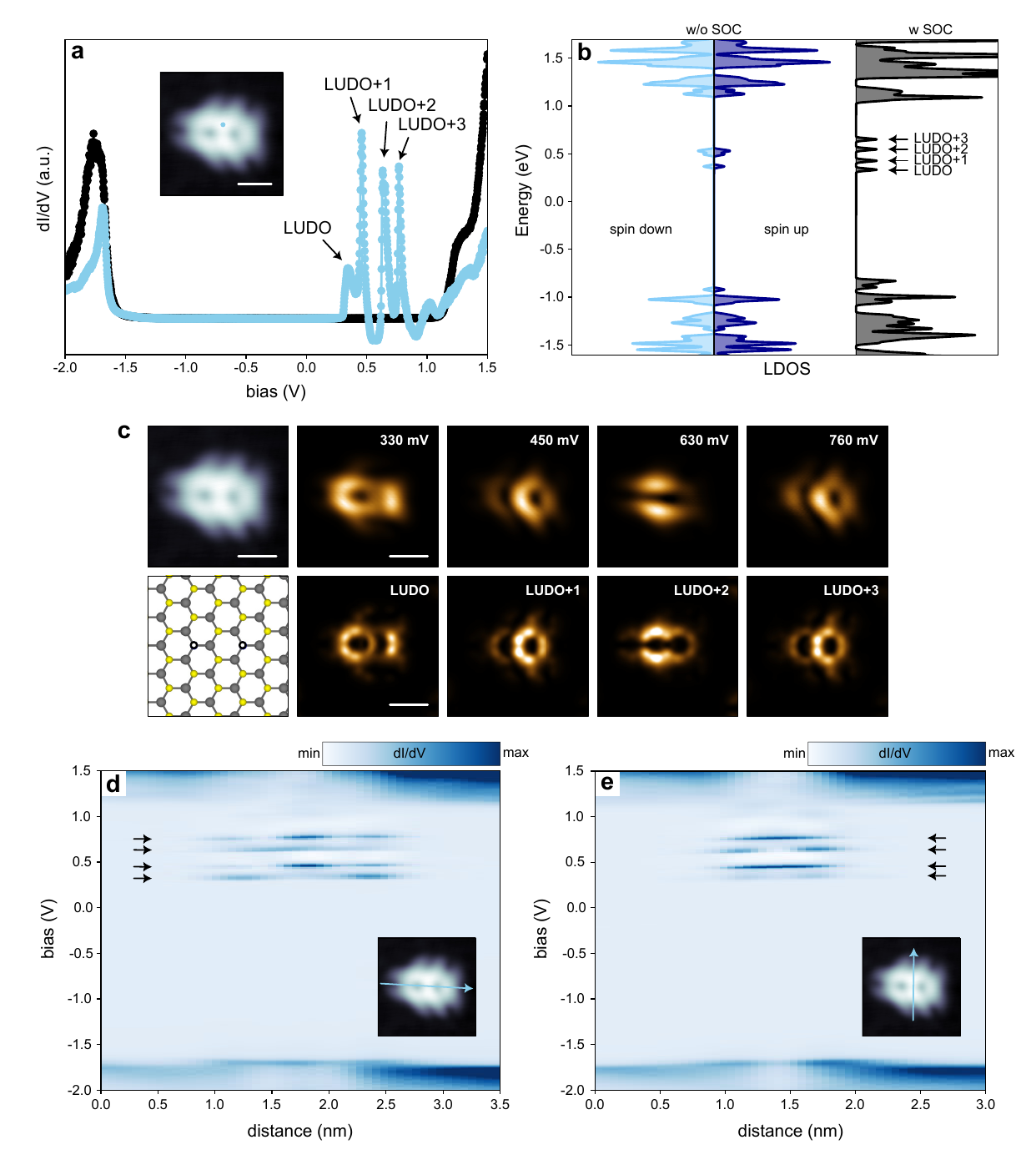}
	\caption{\textbf{Electronic characterization of a sulfur vacancy dimer in a $\sqrt{3}a$-configuration.} \textbf{a} Differential conductance spectra (U$_{\text{bias}}=1.5\, \mathrm{V}$, I$_{\text{set}}=1\, \mathrm{nA}$, U$_{\text{mod}}=5\, \mathrm{mV}$) measured on the dimer lobes (blue) at the position indicated in the inset STM image (U$_{\text{bias}}=1.0\, \mathrm{V}$, I$_{\text{set}}=10\, \mathrm{pA}$, scale bar: $1.0\, \mathrm{nm}$) and on pristine $\mathrm{WS_{2}}$ (black). \textbf{b} Computed LDOS of the dimer obtained by DFT. Spin-polarized LDOS excluding SOC (left) and total LDOS including SOC (right). \textbf{c} Comparison of differential conductance maps measured in constant height mode at the indicated bias voltages (U$_{\text{mod}}=20\, \mathrm{mV}$, scale bar: $1.0\, \mathrm{nm}$) and DFT simulations (scale bar: $1.0\, \mathrm{nm}$) of the four in-gap states. \textbf{d} and \textbf{e} Differential conductance line scans (U$_{\text{mod}}=10\, \mathrm{mV}$) acquired along horizontal and vertical lines across the vacancy dimer respectively. Arrows mark the positions of in-gap states.}
	\label{figure sqrt(3)a-dimer}
\end{figure*}
}

\clearpage

\suppfig{\textbf{Electronic characterization of a sulfur vacancy dimer in a $2a$-configuration}}{
\begin{figure*}[h!]
	\centering
		\includegraphics[width=0.8\textwidth]{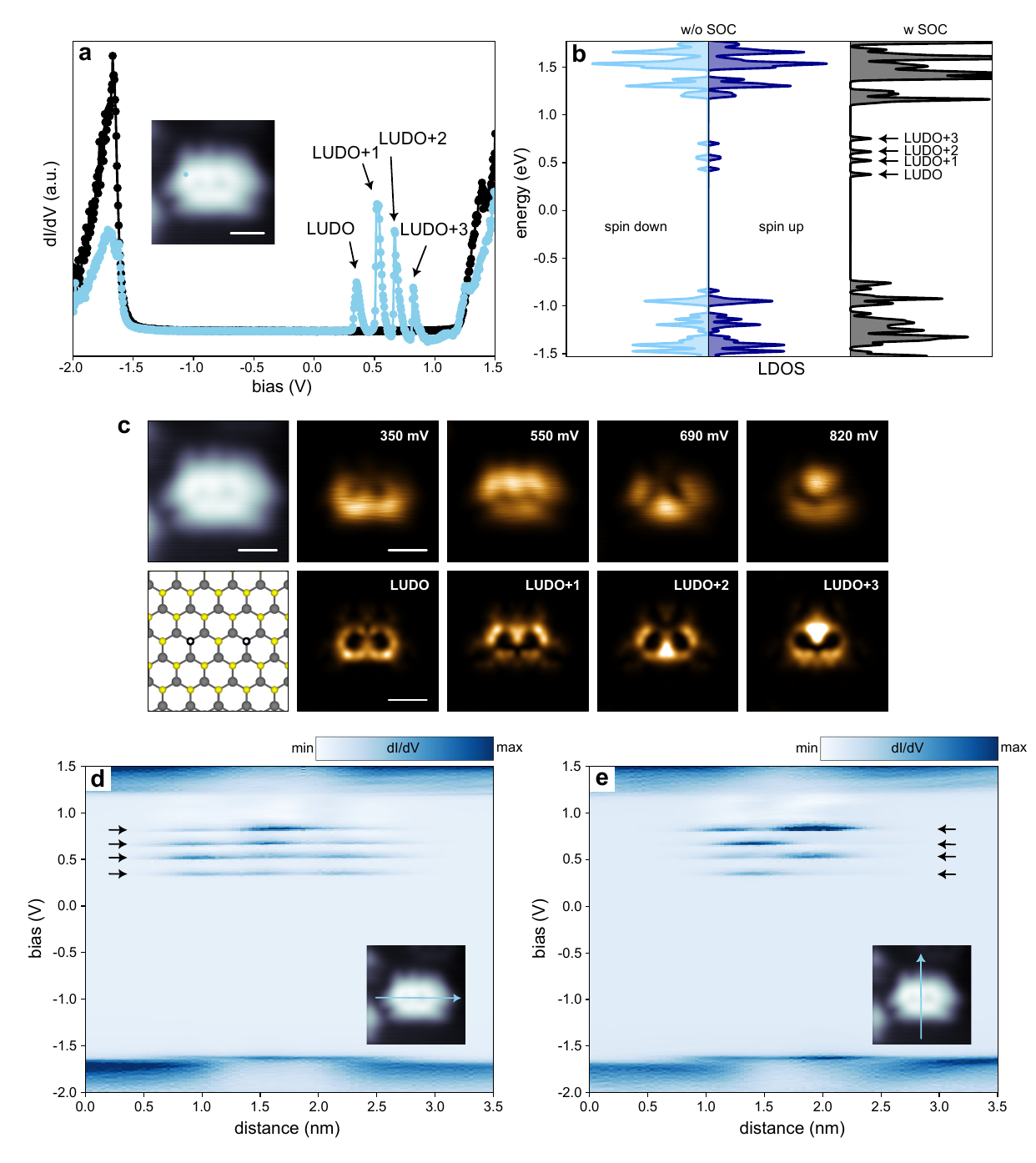}
	\caption{\textbf{Electronic characterization of a sulfur vacancy dimer in a $2a$-configuration.} \textbf{a} Differential conductance spectra (U$_{\text{bias}}=1.5\, \mathrm{V}$, I$_{\text{set}}=1\, \mathrm{nA}$, U$_{\text{mod}}=5\, \mathrm{mV}$) measured on the dimer lobes (blue) at the position indicated in the inset STM image (U$_{\text{bias}}=600\, \mathrm{mV}$, I$_{\text{set}}=40\, \mathrm{pA}$, scale bar: $1.0\, \mathrm{nm}$) and on pristine $\mathrm{WS_{2}}$ (black). \textbf{b} Computed LDOS of the dimer obtained by DFT. Spin-polarized LDOS excluding SOC (left) and total LDOS including SOC (right). \textbf{c} Comparison of differential conductance maps measured in constant height mode at indicated bias voltages (U$_{\text{mod}}=20\, \mathrm{mV}$, scale bar: $1.0\, \mathrm{nm}$) and DFT simulations (scale bar: $1.0\, \mathrm{nm}$) of the four in-gap states. \textbf{d} and \textbf{e} Differential conductance line scans (U$_{\text{mod}}=10\, \mathrm{mV}$) acquired along horizontal and vertical lines across the vacancy dimer respectively. Arrows mark the positions of in-gap states.}
	\label{figure 2a-dimer}
\end{figure*}
}

\clearpage

\suppfig{\textbf{Electronic characterization of a sulfur vacancy dimer in a $3a$-configuration}}{
\begin{figure*}[h!]
	\centering
		\includegraphics[width=0.8\textwidth]{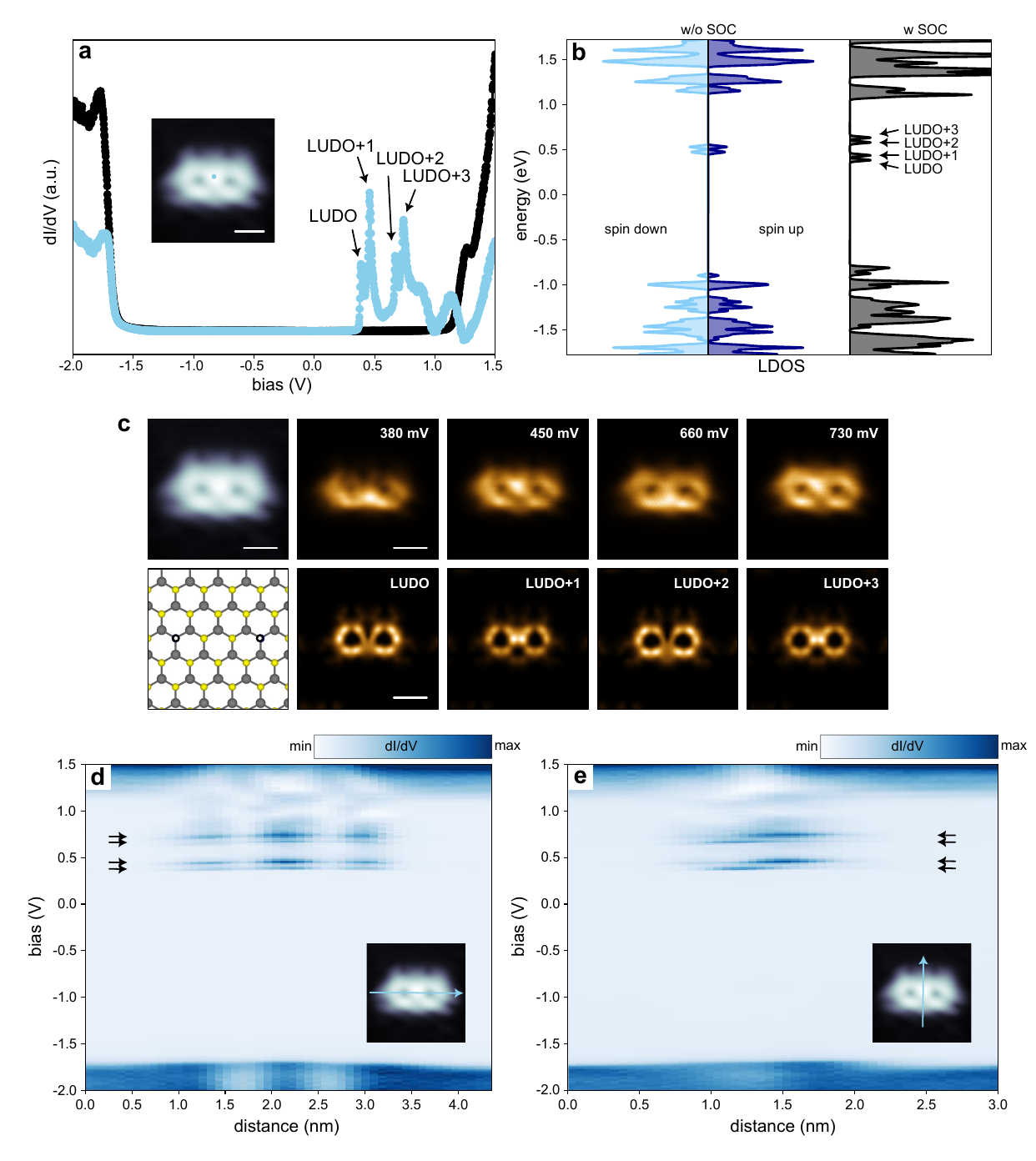}
	\caption{\textbf{Electronic characterization of a sulfur vacancy dimer in a $3a$-configuration.} \textbf{a} Differential conductance spectra (U$_{\text{bias}}=1.5\, \mathrm{V}$, I$_{\text{set}}=1\, \mathrm{nA}$, U$_{\text{mod}}=5\, \mathrm{mV}$) measured on the dimer lobes (blue) at the position indicated in the inset STM image (U$_{\text{bias}}=1.0\, \mathrm{V}$, I$_{\text{set}}=10\, \mathrm{pA}$, scale bar: $1.0\, \mathrm{nm}$) and on pristine $\mathrm{WS_{2}}$ (black). \textbf{b} Computed LDOS of the dimer obtained by DFT. Spin-polarized LDOS excluding SOC (left) and total LDOS including SOC (right). \textbf{c} Comparison of differential conductance maps measured in constant height mode at indicated bias voltages (U$_{\text{mod}}=20\, \mathrm{mV}$, scale bar: $1.0\, \mathrm{nm}$) and DFT simulations (scale bar: $1.0\, \mathrm{nm}$) of the four in-gap states. \textbf{d} and \textbf{e} Differential conductance line scans (U$_{\text{mod}}=10\, \mathrm{mV}$) acquired along horizontal and vertical lines across the vacancy dimer respectively. Arrows mark the positions of in-gap states.}
	\label{figure 3a-dimer}
\end{figure*}
}

\clearpage

\suppfig{\textbf{Rotation of a lateral nearest-neighbor sulfur vacancy dimer caused by tip-induced sulfur diffusion}}{
\begin{figure*}[h!]
	\centering
		\includegraphics[width=1.0\textwidth]{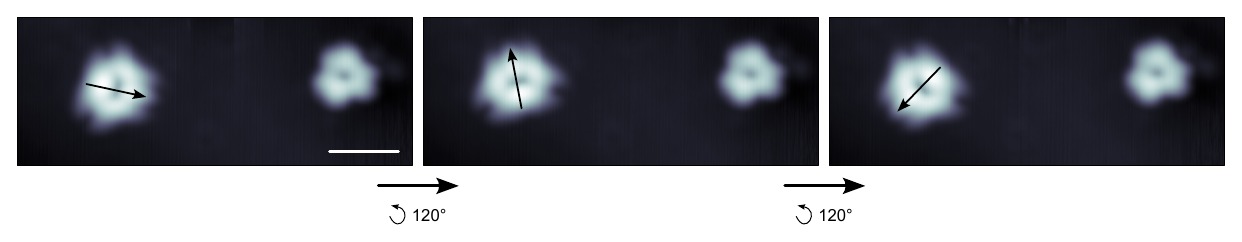}
	\caption{\textbf{Rotation of a lateral nearest-neighbor sulfur vacancy dimer caused by tip-induced sulfur diffusion.} Sequence of STM images (U$_{\text{bias}}=1.0\, \mathrm{V}$, I$_{\text{set}}=5\, \mathrm{pA}$, scale bar: $2.0\, \mathrm{nm}$) showing the three orientational states of the vacancy dimer next to a stationary reference defect.}
	\label{figure 1a-dimer rotation}
\end{figure*}
}

\clearpage

\suppnote{\textbf{Telegraph noise analysis}}{
\noindent For the analysis of the telegraph noise we use the data measured in constant current mode, i.e. with activated feedback loop. Instead of switching between different current states, we then obtain switching between different height states as shown in Fig.~\ref{figure telegraph noise analysis}\textbf{a}. If we look at the current channel of the same data file, we get a plot as shown in Fig.~\ref{figure telegraph noise analysis}\textbf{b}. Each switching event causes a peak in the current signal, as the feedback needs time to react to the sudden change in current. By detecting the peaks, we therefore obtain information about the times at which switching takes place. We analyze the current signal with activated feedback instead of the telegraph noise signals in the current or height channels because the three telegraph noise levels become increasingly difficult to identify with higher frequencies, as separations between them decrease. In order to reduce human bias, we automatized the peak detection. For each telegraph noise trace, we determine a window around the current setpoint covering the noise floor. Only data points above or below that window are considered in the detection. In the plot in Fig.~\ref{figure telegraph noise analysis}\textbf{b} the window is shown by horizontal dotted lines. As a second requirement, an algorithm only detects a data point as a peak if the data point is at least $N$ consecutive data points above or below the noise floor. This number can be set individually for each spectrum to avoid false peak detections close to the noise floor. In Fig.~\ref{figure telegraph noise analysis}\textbf{b} the detected peak data points are marked in red.
After the peaks are detected, the algorithm performs a binning analysis where the number of bins is given by $N_{bins}=\sqrt{N_{events}}$. This part of the analysis has been adopted from Ref.~\citeSM{stipe1998inducing_merged}, where similar telegraph noise investigations are performed. The result is a histogram with the bins on the x-axis and the number of events per bin on the y-axis. As the sulfur diffusion process is induced by inelastic tunneling, the switching statistics follows a Poissonian distribution and the histogram data can be described by
\begin{equation}
    p(\Delta t)=A\cdot\exp{(-\frac{\Delta t}{\tau})=A\cdot\exp{(-R \Delta t)}}
\end{equation}
\noindent with lifetime or dwell time $\tau$ and switching rate $R$. The values can be extracted by a weighted least square fit to the histogram data. The weights $w$ for the fit are determined by
\begin{equation}
    w=\frac{1}{\sigma_{ln(p)}}
\end{equation}
\noindent where $\sigma_{ln(p)}$ is the standard deviation of the logarithmic y-values. Since we are dealing with Poisson statistics, we have $\sigma_{p}=\sqrt{p}$ and therefore
\begin{equation}
    \sigma_{ln(p)} \approx \left|\frac{d(ln(p))}{dp}\right|\cdot\sigma_{p}=\frac{1}{p}\sqrt{p}=\frac{1}{\sqrt{p}}.
\end{equation}
\noindent The fit weight is thus given by the number of counts in each bin:
\begin{equation}
    w=p.
\end{equation}
\noindent An example histogram including the weighted fit for the data in Fig.~\ref{figure telegraph noise analysis}\textbf{a},\textbf{b} is presented in \textbf{c}.
This procedure is repeated for all data files measured at different pairs of bias voltage and tunneling current, providing a data point and its error for each of those files. As shown in Fig.~3\textbf{c} of the main manuscript, the data points belonging to the same bias voltage lie on a line in the double-logarithmic $R(I)$-plot. The slope of the linear functions is determined by a weighted linear fit and plotted as a function of bias voltage in Fig.~3\textbf{d} of the main manuscript. As discussed in detail there, the slope can be interpreted as the number of electrons $N$ that is on average required to induce the diffusion event at the given bias voltage. The value of the CBM from which the second x-axis (upper axis, bias-CBM) in Fig.~3\textbf{d} is derived, is determined from spectra measured on pristine $\mathrm{WS_{2}}$. An example is shown in Fig.~\ref{figure CBM determination}. A statistical analysis of several spectra yields a VBM position of $-1.40\pm0.06\, \mathrm{V}$ and a CBM position of $1.01\pm0.02\, \mathrm{V}$. The gap size of $\mathrm{WS_{2}}$ on graphene/$\mathrm{Ir}$(111) can be quantified as $2.42\pm0.06\, \mathrm{eV}$.
}

\clearpage

\suppfig{\textbf{Telegraph noise analysis}}{
\begin{figure*}[h!]
	\centering
		\includegraphics[width=1.0\textwidth]{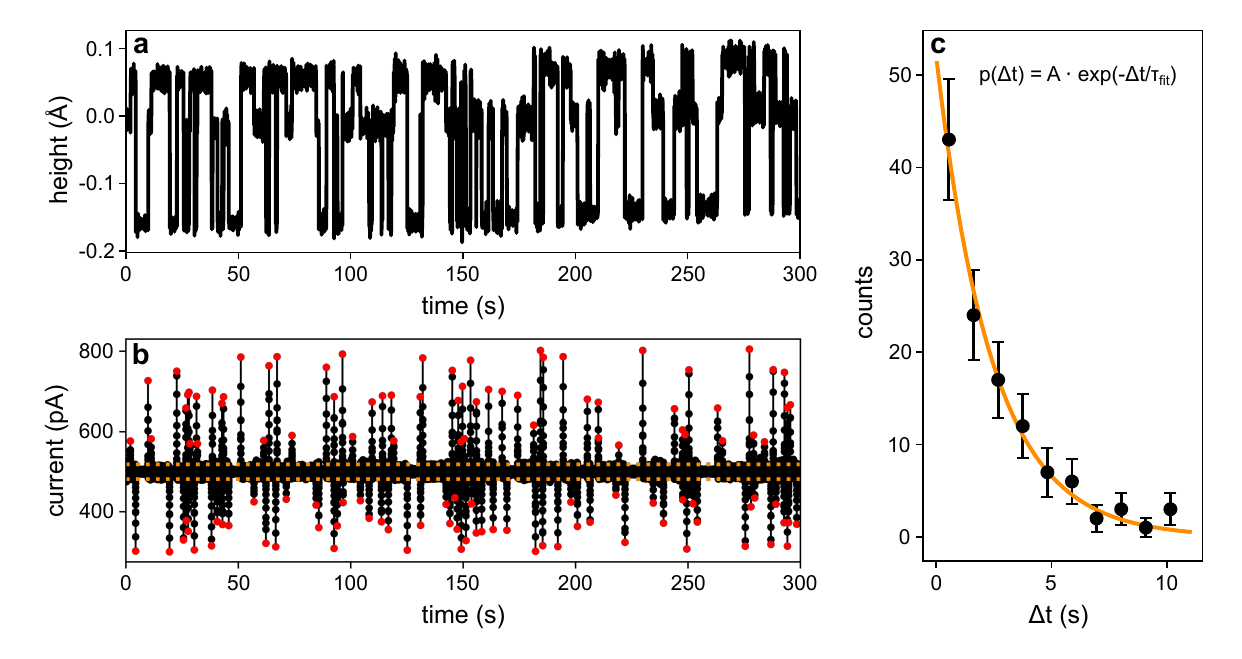}
	\caption{\textbf{Telegraph noise analysis.} \textbf{a} Example of telegraph noise data acquired in constant current mode (feedback activated) at $1.6\, \mathrm{V}$ and $500\, \mathrm{pA}$, shown in the height channel where switching between the three states is visible. \textbf{b} Same measurement as in \textbf{a} shown in the current channel. Each peak corresponds to a switching event. Peak detection is performed automatically, see text. Noise floor around current setpoint indicated by orange dotted lines. We define a switching event when at least four consecutive data points are above or below the noise floor. These events are identified as peaks and marked in red. \textbf{c} Histogram resulting from the peak detection in \textbf{b} and Poissonian fit to the data (orange curve) with fit parameters $A=52.22$ and $\tau_{\text{fit}}=2.42\pm0.42\, \mathrm{s}$.}
	\label{figure telegraph noise analysis}
\end{figure*}
}

\clearpage

\suppfig{\textbf{Determination of $\mathrm{WS_{2}}$ band edges}}{
\begin{figure*}[h!]
	\centering
		\includegraphics[width=1.0\textwidth]{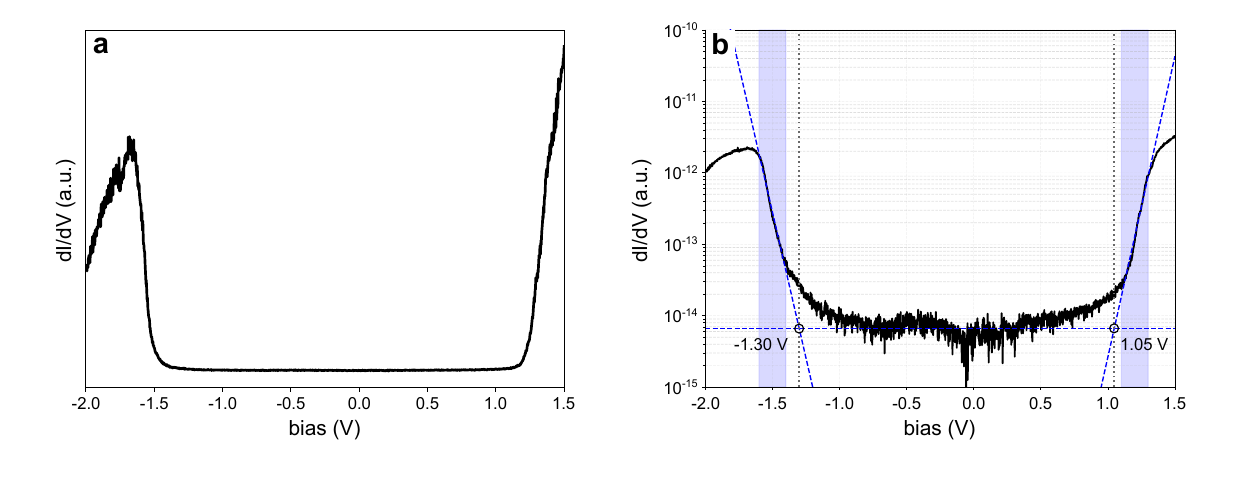}
	\caption{\textbf{Determination of $\mathrm{WS_{2}}$ band edges.} \textbf{a} Differential conductance spectrum (U$_{\text{bias}}=1.5\, \mathrm{V}$, I$_{\text{set}}=100\, \mathrm{pA}$, U$_{\text{mod}}=5\, \mathrm{mV}$) measured in constant height mode on a pristine $\mathrm{WS_{2}}$ area. \textbf{b} Determination of band edge positions from the spectrum shown in \textbf{a} plotted on a logarithmic scale. The horizontal dotted blue line is the zero-level of the band gap. The other two dotted blue lines are linear functions fitted to the steep parts of the band edges which are marked by the shaded blue regions. The valence and conduction band edges are then determined by the intersection of the linear fits with the gap zero-line and are given by $-1.3\, \mathrm{V}$ and $1.05\, \mathrm{V}$, respectively. The gap size is therefore $2.35\, \mathrm{eV}$.}
	\label{figure CBM determination}
\end{figure*}
}

\clearpage

\suppnote{\textbf{Sulfur diffusion barrier calculations}}{
\noindent The diffusion barrier for a sulfur atom next to two sulfur vacancies is shown in Fig.~3\textbf{e} in the main manuscript. The diffusion path via the hollow site has the lowest energy barrier of $1.19\, \mathrm{eV}$, which is in good agreement with previous calculations~\citeSM{wang2024comparative_merged} and the experimentally estimated value. We also considered if the agreement could be improved further by including additional effects. First, we considered whether the electric field between the tip and the sample could modify the diffusion barrier. We modeled this by adding a homogeneous out-of-plane electric field to the calculations. However, this had virtually no effect on the barrier. Second, we considered the role of nuclear quantum corrections to the lowest order by evaluating the zero-point energy (ZPE) correction to the diffusion barrier~\citeSM{milas2011diffusion_merged}:

\begin{equation}
    E^{\text{ZPE}} = E^{\text{classical}}+\frac{1}{2} \sum_{i \in \text{TS}} \hbar \omega_{i}-\frac{1}{2} \sum_{j \in \text{IS}} \hbar \omega_{j}
\end{equation}

\noindent where IS denotes initial states and TS transitional states. Only real vibrational modes are included in the sums and since the transition state has one fewer real vibrational modes (one being imaginary), it generally leads to lower barrier. We carried these calculations at smaller $4\times4$ supercells to reduce the number of displacement calculations. Cutoff was set at $350\, \mathrm{eV}$ and the k-point mesh at $3\times3$. The calculated diffusion barrier at these settings remained at $1.19\, \mathrm{eV}$. We obtained a ZPE correction of $-40.03\, \mathrm{meV}$, which is comparable to typical phonon energies, and consequently a corrected diffusion barrier of $1.15\, \mathrm{eV}$. The uncorrected and ZPE-corrected diffusion pathways are plotted in Fig.~\ref{figure ZPE-corrected diffusion barrier}.
}

\clearpage

\suppfig{\textbf{Sulfur diffusion barrier calculations corrected for zero-point energies}}{
\begin{figure*}[h!]
	\centering
		\includegraphics[width=1.0\textwidth]{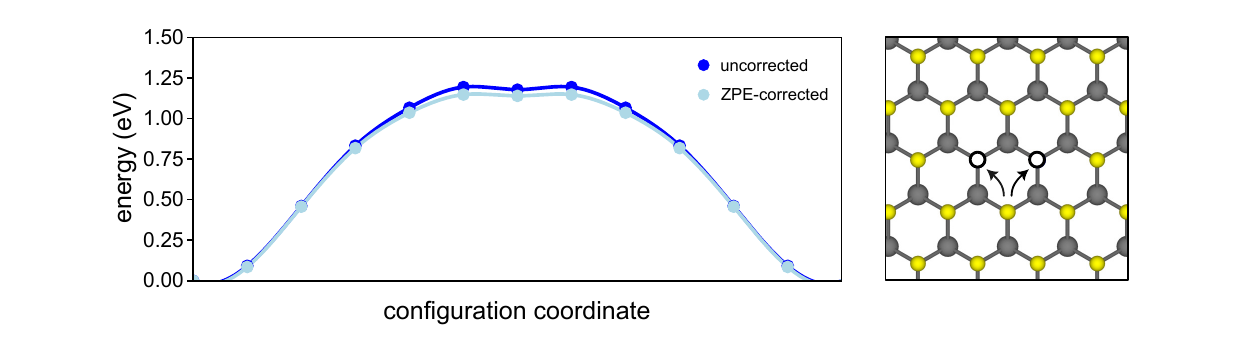}
	\caption{\textbf{Sulfur diffusion barrier calculations corrected for zero-point energies.} Comparison between the uncorrected (blue) and ZPE-corrected (light blue) diffusion paths. The ZPE correction leads to a barrier reduction of $40.03\, \mathrm{meV}$. The calculations are done for sulfur diffusion via the hollow-site in presence of a lateral nearest-neighbor sulfur vacancy dimer. Sulfur vacancy positions (open black circles) and equivalent diffusion paths via hollow sites (black arrows) are shown in the ball model.}
	\label{figure ZPE-corrected diffusion barrier}
\end{figure*}
}

\clearpage

\suppfig{\textbf{Electronic characterization of a straight sulfur vacancy trimer line}}{
\begin{figure*}[h!]
	\centering
		\includegraphics[width=0.8\textwidth]{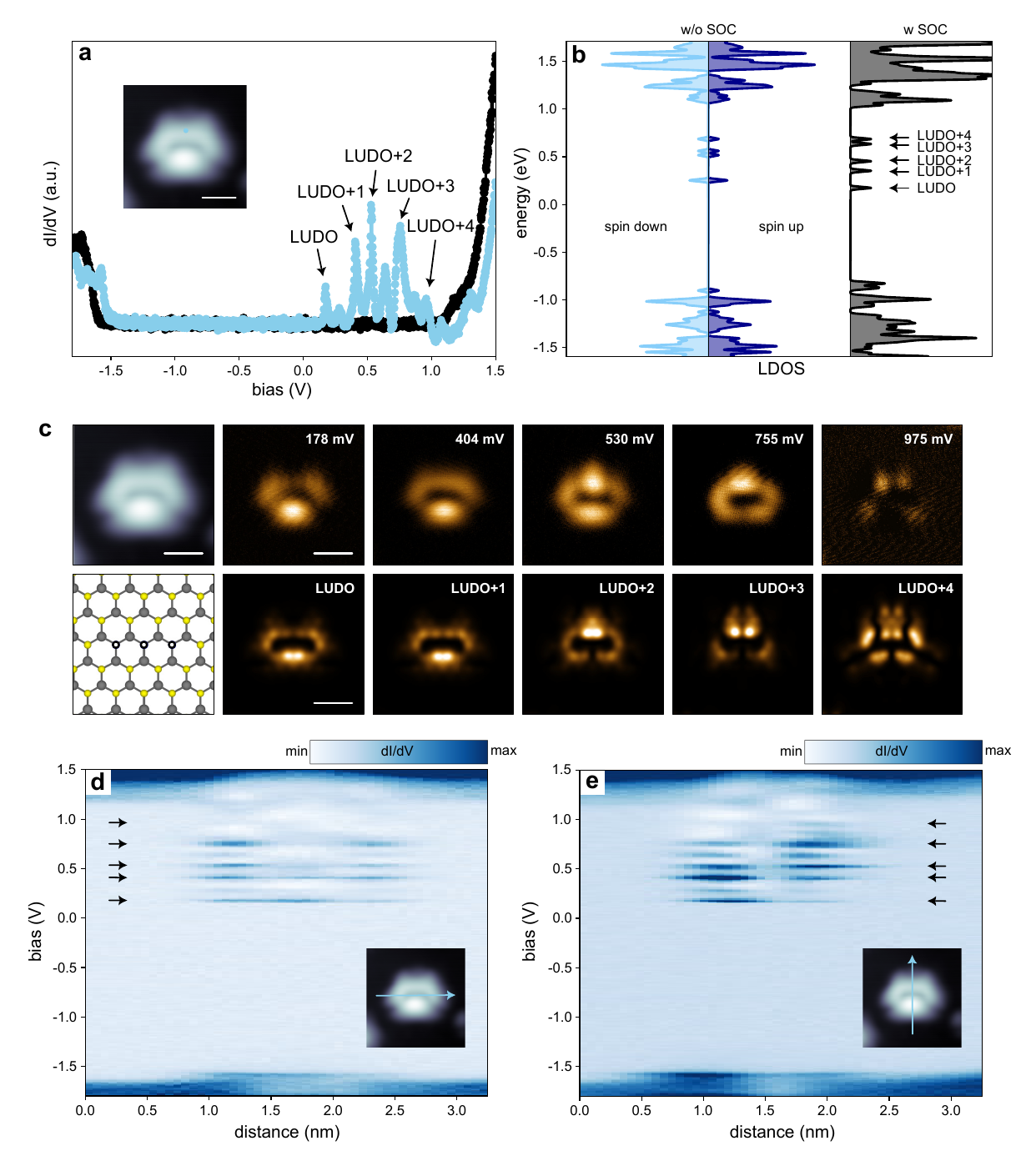}
	\caption{\textbf{Electronic characterization of a straight sulfur vacancy trimer line.} \textbf{a} Differential conductance spectra (U$_{\text{bias}}=1.5\, \mathrm{V}$, I$_{\text{set}}=100\, \mathrm{pA}$, U$_{\text{mod}}=5\, \mathrm{mV}$) measured on the lobes of the trimer (blue) at the position indicated in the inset STM image (U$_{\text{bias}}=800\, \mathrm{mV}$, I$_{\text{set}}=40\, \mathrm{pA}$, scale bar: $1.0\, \mathrm{nm}$) and on pristine $\mathrm{WS_{2}}$ (black). \textbf{b} Computed LDOS of the trimer obtained by DFT. Spin-polarized LDOS excluding SOC (left) and total LDOS including SOC (right). \textbf{c} Comparison of differential conductance maps measured in constant height mode at indicated bias voltages (U$_{\text{mod}}=20\, \mathrm{mV}$, scale bar: $1.0\, \mathrm{nm}$) and DFT simulations (scale bar: $1.0\, \mathrm{nm}$) of the five in-gap states. \textbf{d} and \textbf{e} Differential conductance line scans (U$_{\text{mod}}=10\, \mathrm{mV}$) acquired along horizontal and vertical lines across the vacancy trimer respectively. Arrows mark the positions of in-gap states.}
	\label{figure straight 1a-line}
\end{figure*}
}

\clearpage

\suppfig{\textbf{Electronic characterization of a kinked line sulfur vacancy trimer}}{
\begin{figure*}[h!]
	\centering
		\includegraphics[width=0.8\textwidth]{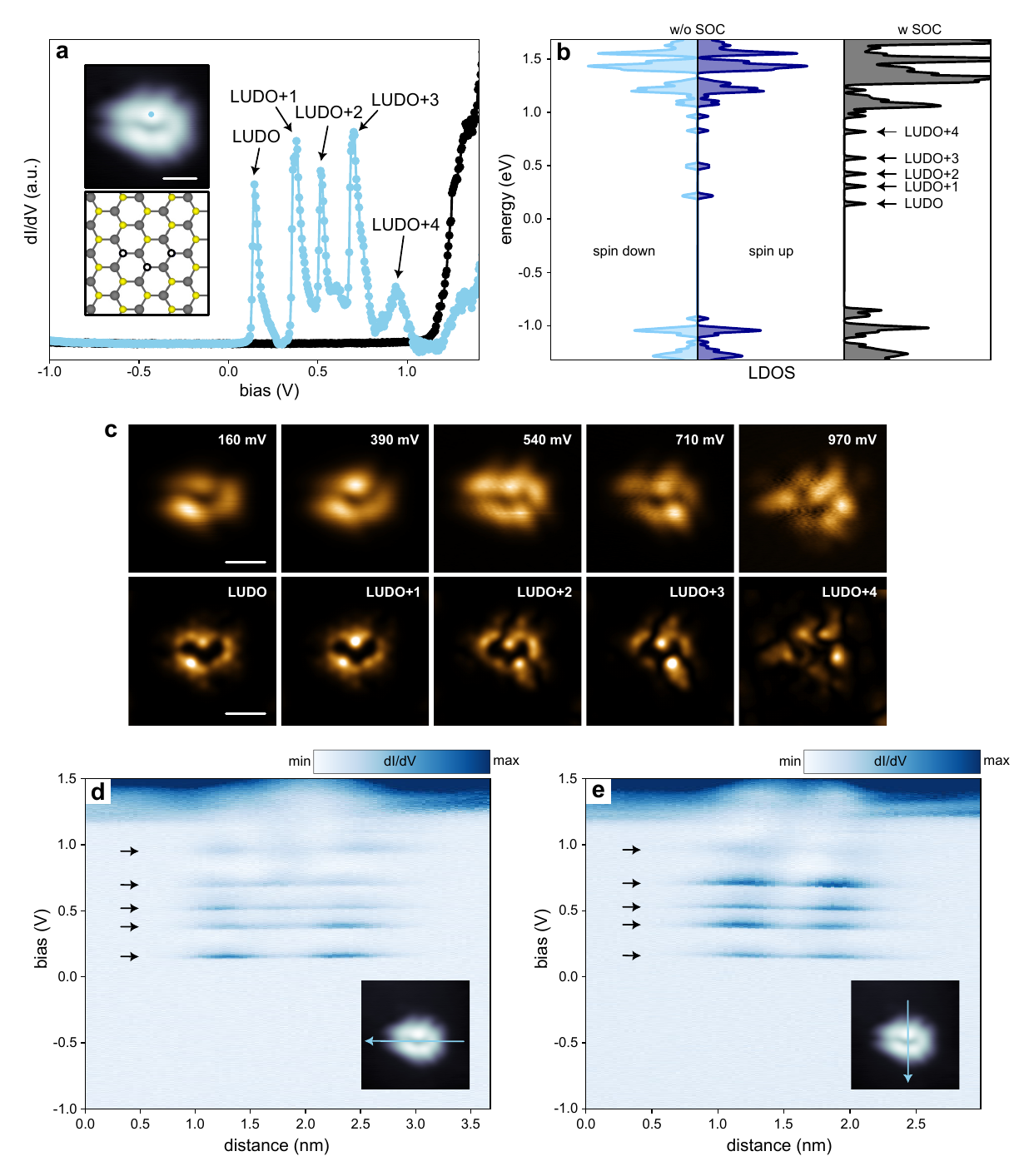}
	\caption{\textbf{Electronic characterization of a kinked line sulfur vacancy trimer.} \textbf{a} Differential conductance spectra (U$_{\text{bias}}=1.5\, \mathrm{V}$, I$_{\text{set}}=100\, \mathrm{pA}$, U$_{\text{mod}}=5\, \mathrm{mV}$) measured on the lobes of the trimer (blue) at the position indicated in the inset STM image (U$_{\text{bias}}=1.0\, \mathrm{V}$, I$_{\text{set}}=5\, \mathrm{pA}$, scale bar: $1.0\, \mathrm{nm}$) and on pristine $\mathrm{WS_{2}}$ (black). \textbf{b} Computed LDOS of the trimer obtained by DFT. Spin-polarized LDOS excluding SOC (left) and total LDOS including SOC (right). \textbf{c} Comparison of differential conductance maps measured in constant height mode at indicated bias voltages (U$_{\text{mod}}=20\, \mathrm{mV}$, scale bar: $1.0\, \mathrm{nm}$) and DFT simulations (scale bar: $1.0\, \mathrm{nm}$) of the five in-gap states. \textbf{d} and \textbf{e} Differential conductance line scans (U$_{\text{mod}}=5\, \mathrm{mV}$) acquired along horizontal and vertical lines across the vacancy trimer respectively. Arrows mark the positions of in-gap states.}
	\label{figure kinked 1a-line}
\end{figure*}
}

\clearpage

\suppfig{\textbf{Optical transition dipole maps of the hollow-centered triangle trimer.}}{
\begin{figure*}[h!]
	\centering
		\includegraphics[width=0.8\textwidth]{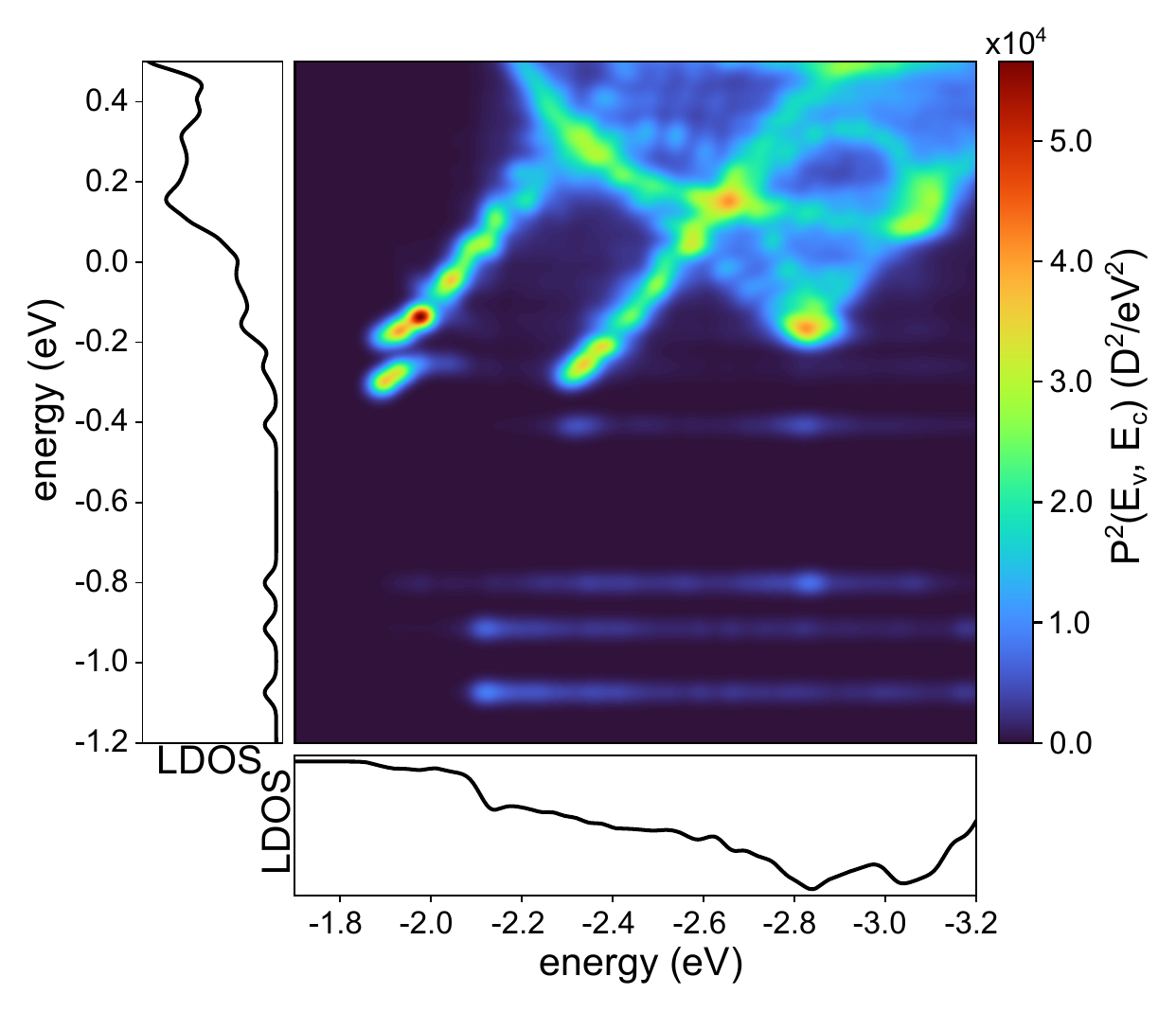}
	\caption{\textbf{Optical transition dipole map of the hollow-centered triangle trimer.}}
	\label{figure dipole map VS3(tri-hole)}
\end{figure*}
}

\clearpage

\suppfig{\textbf{Straight line sulfur vacancy tetramer}}{
\begin{figure*}[h!]
	\centering
		\includegraphics[width=1.0\textwidth]{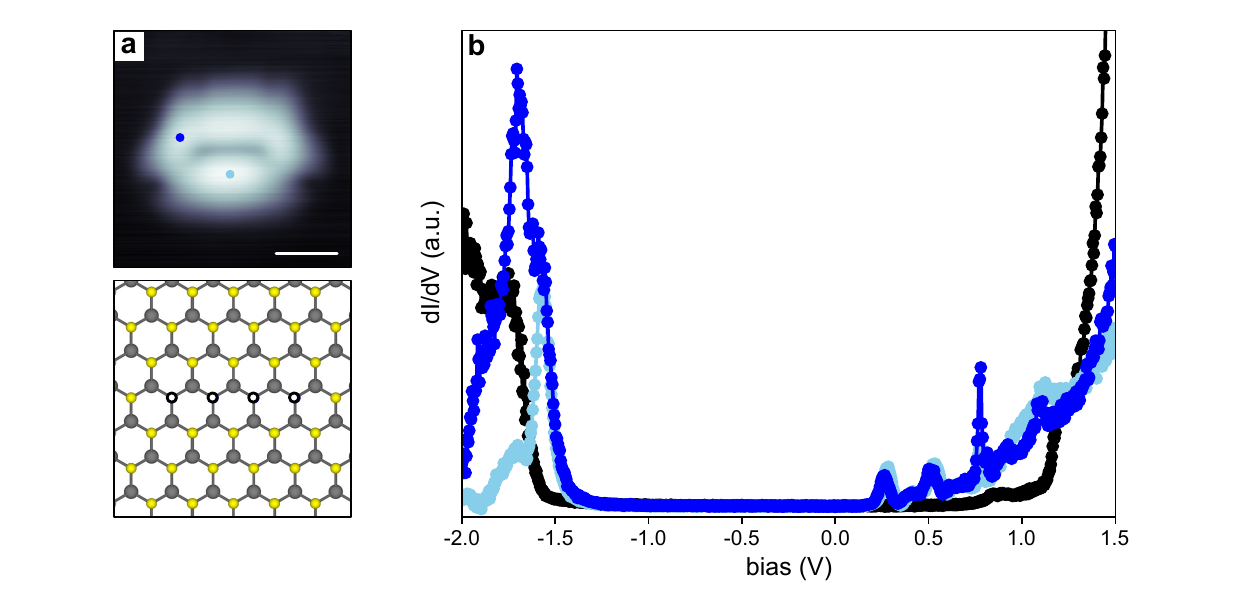}
	\caption{\textbf{Straight line sulfur vacancy tetramer.} \textbf{a} STM image (U$_{\text{bias}}=1.0\, \mathrm{V}$, I$_{\text{set}}=40\, \mathrm{pA}$, scale bar: $1.0\, \mathrm{nm}$) and ball model of the sulfur vacancy configuration. \textbf{b} Differential conductance spectra (U$_{\text{bias}}=1.5\, \mathrm{V}$, I$_{\text{set}}=800\, \mathrm{pA}$, U$_{\text{mod}}=5\, \mathrm{mV}$) measured on the defect lobes at positions indicated in \textbf{a} (light and dark blue) as well as on prisitine $\mathrm{WS_{2}}$ (black).}
	\label{figure 1a-tetramer}
\end{figure*}
}

\clearpage

\suppfig{\textbf{Electronic characterization of a V-shaped sulfur vacancy tetramer}}{
\begin{figure*}[h!]
	\centering
		\includegraphics[width=1.0\textwidth]{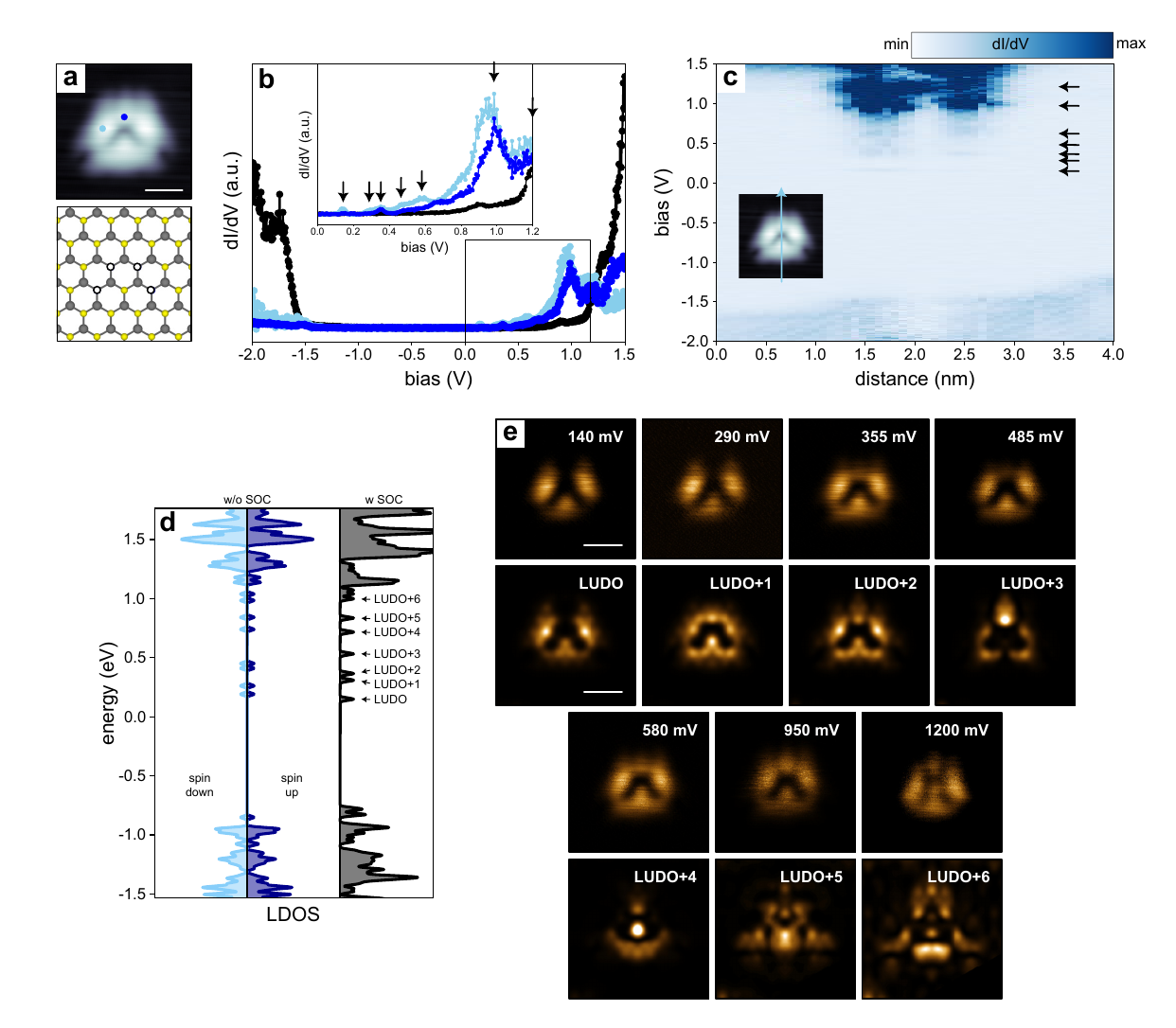}
	\caption{\textbf{Electronic characterization of a V-shaped sulfur vacancy tetramer.} \textbf{a} STM image (U$_{\text{bias}}=1.0\, \mathrm{V}$, I$_{\text{set}}=40\, \mathrm{pA}$, scale bar: $1.0\, \mathrm{nm}$) and ball model of a V-shaped sulfur vacancy tetramer. \textbf{b} Differential conductance spectra (U$_{\text{bias}}=1.5\, \mathrm{V}$, I$_{\text{set}}=500\, \mathrm{pA}$, U$_{\text{mod}}=5\, \mathrm{mV}$) measured on the tetramer lobes (light and dark blue) at positions indicated in \textbf{a} as well as on pristine $\mathrm{WS_{2}}$ (black). \textbf{c} Differential conductance line scan (U$_{\text{mod}}=10\, \mathrm{mV}$) measured along a line over the cluster as depicted in the inset. \textbf{d} Computed LDOS of the tetramer obtained by DFT. Spin-polarized LDOS excluding SOC (left) and total LDOS including SOC (right). \textbf{e} Differential conductance maps (U$_{\text{mod}}=20\, \mathrm{mV}$, scale bar: $1.0\, \mathrm{nm}$) measured in constant height mode at indicated bias voltages. The bias voltages of the maps are also highlighted by the arrows in \textbf{c} and \textbf{d}. Tentative comparison to DFT simulations (scale bar: $1.0\, \mathrm{nm}$) of in-gap states.}
	\label{figure V-shaped tetramer}
\end{figure*}
}

\clearpage

\suppfig{\textbf{Electronic characterization of the left-handed chiral sulfur vacancy tetramer}}{
\begin{figure*}[h!]
	\centering
		\includegraphics[width=1.0\textwidth]{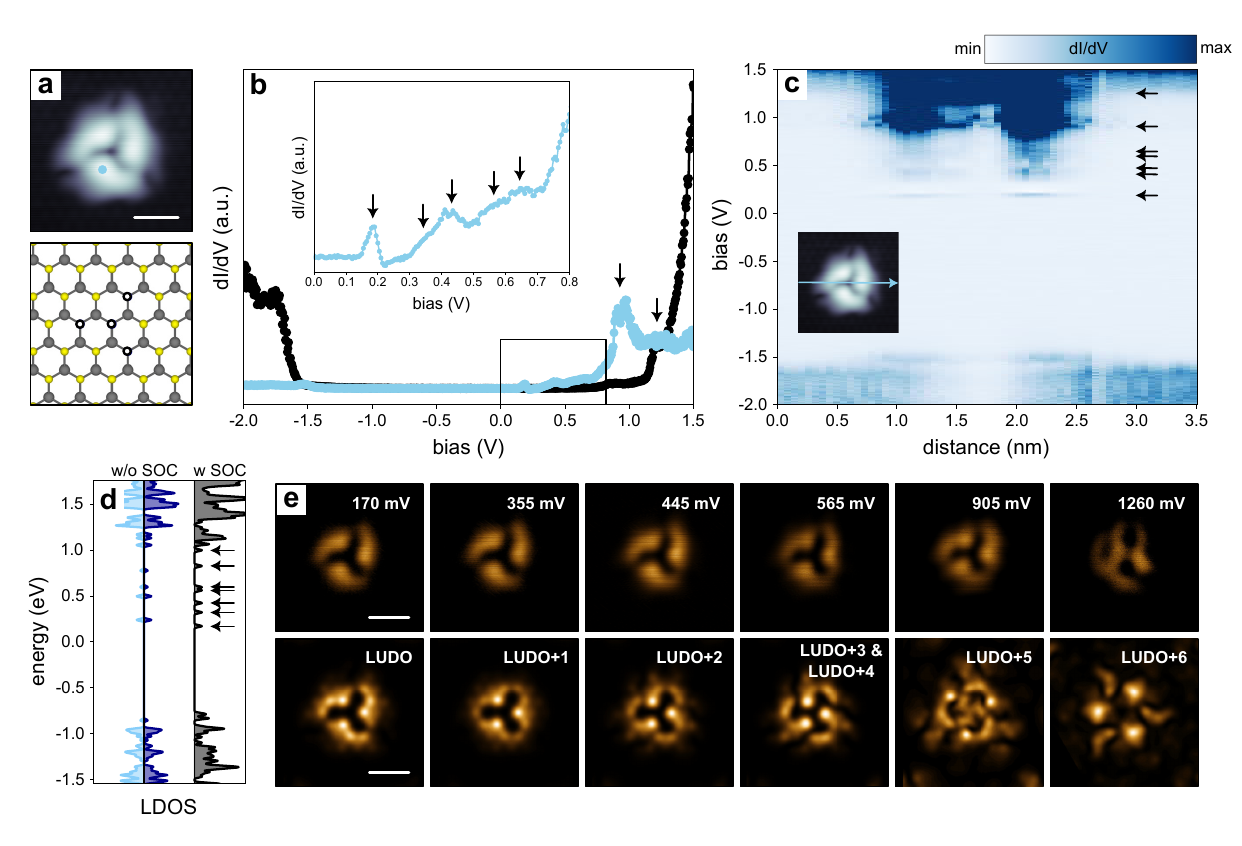}
	\caption{\textbf{Electronic characterization of the left-handed chiral sulfur vacancy tetramer.} \textbf{a} STM image (U$_{\text{bias}}=500\, \mathrm{mV}$, I$_{\text{set}}=40\, \mathrm{pA}$, scale bar: $1.0\, \mathrm{nm}$) and ball model of the tetramer configuration. \textbf{b} Differential conductance spectra (U$_{\text{bias}}=1.5\, \mathrm{V}$, I$_{\text{set}}=800\, \mathrm{pA}$, U$_{\text{mod}}=5\, \mathrm{mV}$) measured on the tetramer lobe marked in \textbf{a} (blue) and on pristine $\mathrm{WS_{2}}$ (black). The inset displays a zoomed region of the spectrum. Peaks associated with in-gap states are highlighted by arrows. \textbf{c} Differential conductance line scan (U$_{\text{mod}}=10\, \mathrm{mV}$) acquired along a line across the tetramer as shown in the inset. \textbf{d} Computed LDOS of the tetramer obtained by DFT. Spin-polarized LDOS excluding SOC (left) and total LDOS including SOC (right). In-gap states of the defect are marked by arrows. \textbf{e} Tentative comparison of differential conductance maps measured in constant height mode at indicated bias voltages (U$_{\text{mod}}=20\, \mathrm{mV}$, scale bar: $1.0\, \mathrm{nm}$) and DFT simulations (scale bar: $1.0\, \mathrm{nm}$) of in-gap states.}
	\label{figure left-handed chiral tetramer}
\end{figure*}
}

\clearpage

\suppfig{\textbf{Electronic characterization of the right-handed chiral sulfur vacancy tetramer}}{
\begin{figure*}[h!]
	\centering
		\includegraphics[width=1.0\textwidth]{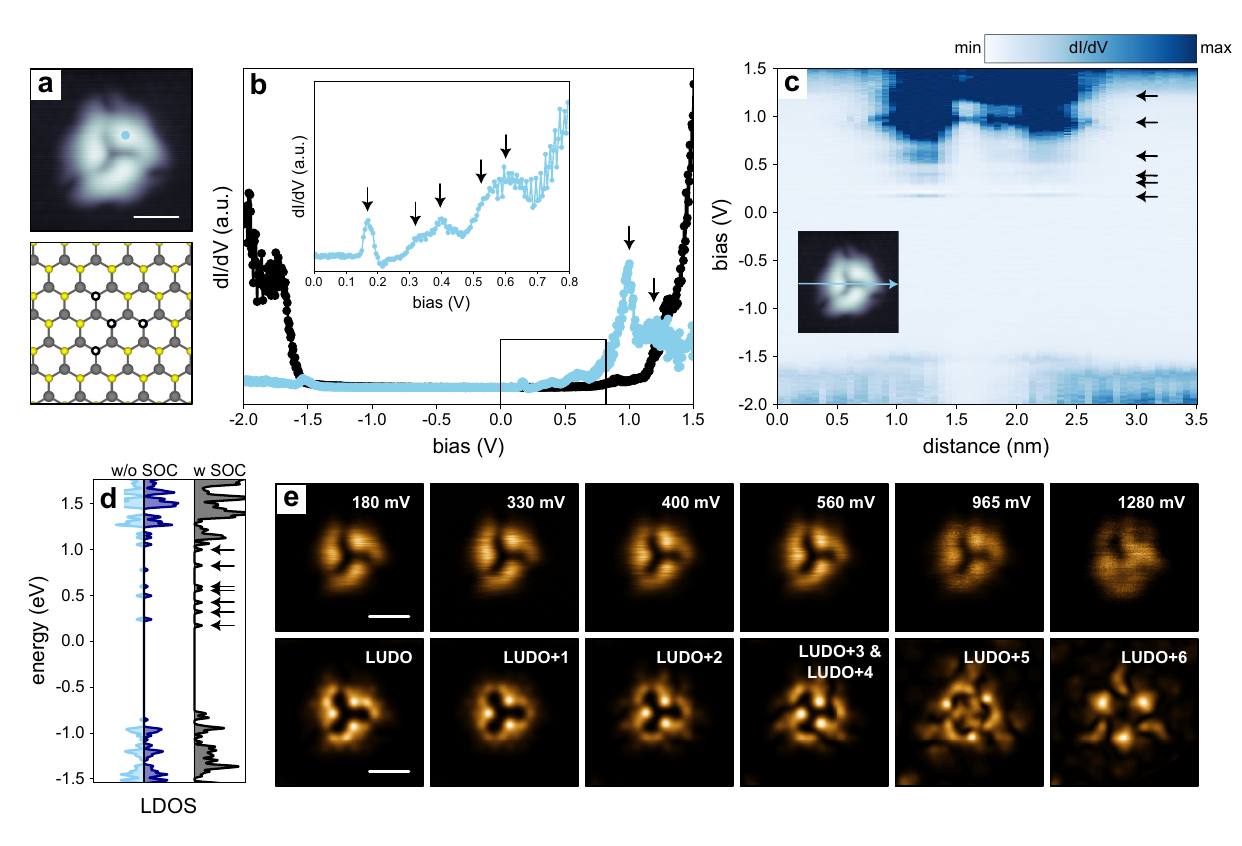}
	\caption{\textbf{Electronic characterization of the right-handed chiral sulfur vacancy tetramer.} \textbf{a} STM image (U$_{\text{bias}}=500\, \mathrm{mV}$, I$_{\text{set}}=40\, \mathrm{pA}$, scale bar: $1.0\, \mathrm{nm}$) and ball model of the tetramer configuration. \textbf{b} Differential conductance spectra (U$_{\text{bias}}=1.5\, \mathrm{V}$, I$_{\text{set}}=800\, \mathrm{pA}$, U$_{\text{mod}}=5\, \mathrm{mV}$) measured on the tetramer lobe marked in \textbf{a} (blue) and on pristine $\mathrm{WS_{2}}$ (black). The inset displays a zoomed region of the spectrum. Peaks associated with in-gap states are highlighted by arrows. \textbf{c} Differential conductance line scan (U$_{\text{mod}}=10\, \mathrm{mV}$) acquired along a line across the tetramer as shown in the inset. \textbf{d} Computed LDOS of the tetramer obtained by DFT. Spin-polarized LDOS excluding SOC (left) and total LDOS including SOC (right). In-gap states of the defect are marked by arrows. \textbf{e} Tentative comparison of differential conductance maps measured in constant height mode at indicated bias voltages (U$_{\text{mod}}=20\, \mathrm{mV}$, scale bar: $1.0\, \mathrm{nm}$) and DFT simulations (scale bar: $1.0\, \mathrm{nm}$) of in-gap states.}
	\label{figure right-handed chiral tetramer}
\end{figure*}
}

\clearpage

\suppnote{\textbf{Energetics}}{
\noindent We calculate the formation $E_{f}$ and binding energies $E_{b}$ of the investigated sulfur vacancy complexes as
\begin{equation}
    E_{\text{f}}[NV_{S}] = E[NV_{S}] - (E[0]-N\mu_{S})
\end{equation}
and
\begin{equation}
    E_{b} = E_{f}[NV_{S}]-NE_{f}[V_{S}] = E[NV_{S}]-E[0]-N(E[V_{S}]-E[0])
\end{equation}
which are listed in Tab.~\ref{tab: binding energies}. Negative binding energies indicate more stable configurations, that is, the 1a-dimer, vacancy lines and the chiral tetramer, in agreement with previous reports~\citeSM{komsa2013defects_merged, Le2014vacancies_merged, flototto2026cooperative_merged, wang2022first_merged, wang2024comparative_merged}.
}
\supptab{\textbf{Binding energies of selected lateral sulfur vacancy complexes}}{
\begin{table}[h]
    \centering
    \setlength{\tabcolsep}{10pt}
    \begin{tabular}{|c|c|c|}
        \hline
        configuration & abbreviation & binding energy ($\mathrm{meV}$) \\
        \hline
        1a-dimer & $2V_{S}(1,0)$ & $-15.4$ \\
        $\sqrt{3}$a-dimer & $2V_{S}(\sqrt{3})$ & $5.8$ \\
        2a-dimer & $2V_{S}(2,0)$ & $41.5$ \\
        3a-dimer & $2V_{S}(3,0)$ & $29.8$ \\
        straight trimer line & $3V_{S}(line)$ & $-111.1$ \\
        kinked trimer line & $3V_{S}(kink)$ & $-30.2$ \\
        hollow-centered triangle trimer & $3V_{S}(tri-hollow)$ & $226.6$ \\
        metal-centered triangle trimer & $3V_{S}(tri-W)$ & $310.2$ \\
        V-shaped tetramer & $4V_{S}(V)$ & $77.2$ \\
        chiral tetramer & $4V_{S}(chiral)$ & $-47.7$ \\
        \hline
    \end{tabular}
    \caption{\textbf{Binding energies of selected lateral sulfur vacancy complexes.}}
    \label{tab: binding energies}
\end{table}
}

\clearpage

\addcontentsline{toc}{section}{References}
\addtocontents{toc}{\protect\setcounter{tocdepth}{-1}}

\addcontentsline{toc}{section}{References}

\begingroup
\renewcommand{\addcontentsline}[3]{}   
\section*{References}
\makeatletter
\@ifundefined{c@NAT@ctr}{}{\setcounter{NAT@ctr}{0}}
\setcounter{enumiv}{0}
\makeatother
\bibliographystyleSM{apsrev4-2}
\bibliographySM{./library_SM_merged}
\endgroup

\end{document}